\documentclass[aps,english,preprint,longbibliography]{revtex4-1}
\usepackage{amsmath}
\usepackage{SIunits}
\usepackage{graphicx}
\usepackage{bbold}

\begin{document}
%
%
\newcommand{\ac}[0]{\ensuremath{\hat{a}_{\mathrm{c}}}}
\newcommand{\adagc}[0]{\ensuremath{\hat{a}^{\dagger}_{\mathrm{c}}}}
\newcommand{\aR}[0]{\ensuremath{\hat{a}_{\mathrm{R}}}}
\newcommand{\aT}[0]{\ensuremath{\hat{a}_{\mathrm{T}}}}
\renewcommand{\b}[0]{\ensuremath{\hat{b}}}
\newcommand{\bdag}[0]{\ensuremath{\hat{b}^{\dagger}}}
\newcommand{\betaI}[0]{\ensuremath{\beta_\mathrm{I}}}
\newcommand{\betaR}[0]{\ensuremath{\beta_\mathrm{R}}}
\renewcommand{\c}[0]{\ensuremath{\hat{c}}}
\newcommand{\cdag}[0]{\ensuremath{\hat{c}^{\dagger}}}
\newcommand{\CorrMat}[0]{\ensuremath{\boldsymbol\gamma}}
\newcommand{\Deltacs}[0]{\ensuremath{\Delta_{\mathrm{cs}}}}
\newcommand{\Deltacsmax}[0]{\ensuremath{\Delta_{\mathrm{cs}}^{\mathrm{max}}}}
\newcommand{\Deltacsparked}[0]{\ensuremath{\Delta_{\mathrm{cs}}^{\mathrm{p}}}}
\newcommand{\Deltacstarget}[0]{\ensuremath{\Delta_{\mathrm{cs}}^{\mathrm{t}}}}
\newcommand{\Deltae}[0]{\ensuremath{\Delta_{\mathrm{e}}}}
\newcommand{\Deltahfs}[0]{\ensuremath{\Delta_{\mathrm{hfs}}}}
\newcommand{\dens}[0]{\ensuremath{\hat{\rho}}}
\newcommand{\erfc}[0]{\ensuremath{\mathrm{erfc}}}
\newcommand{\Fq}[0]{\ensuremath{F_{\mathrm{q}}}}
\newcommand{\gammapar}[0]{\ensuremath{\gamma_{\parallel}}}
\newcommand{\gammaperp}[0]{\ensuremath{\gamma_{\perp}}}
\newcommand{\gavg}[0]{\ensuremath{\mathcal{G}_{\mathrm{avg}}}}
\newcommand{\gbar}[0]{\ensuremath{\bar{g}}}
\newcommand{\gens}[0]{\ensuremath{g_{\mathrm{ens}}}}
\renewcommand{\H}[0]{\ensuremath{\hat{H}}}
\renewcommand{\Im}[0]{\ensuremath{\mathrm{Im}}}
\newcommand{\kappac}[0]{\ensuremath{\kappa_{\mathrm{c}}}}
\newcommand{\kappamin}[0]{\ensuremath{\kappa_{\mathrm{min}}}}
\newcommand{\kappamax}[0]{\ensuremath{\kappa_{\mathrm{max}}}}
\newcommand{\ket}[1]{\ensuremath{|#1\rangle}}
\newcommand{\mat}[1]{\ensuremath{\mathbf{#1}}}
\newcommand{\mean}[1]{\ensuremath{\langle#1\rangle}}
\newcommand{\omegac}[0]{\ensuremath{\omega_{\mathrm{c}}}}
\newcommand{\omegas}[0]{\ensuremath{\omega_{\mathrm{s}}}}
\newcommand{\pauli}[0]{\ensuremath{\hat{\sigma}}}
\newcommand{\pexc}[0]{\ensuremath{p_{\mathrm{exc}}}}
\newcommand{\pexceff}[0]{\ensuremath{p_{\mathrm{exc}}^{\mathrm{eff}}}}
\newcommand{\Pa}[0]{\ensuremath{\hat{P}_{\mathrm{c}}}}
\newcommand{\Qmin}[0]{\ensuremath{Q_{\mathrm{min}}}}
\newcommand{\Qmax}[0]{\ensuremath{Q_{\mathrm{max}}}}
\renewcommand{\Re}[0]{\ensuremath{\mathrm{Re}}}
\renewcommand{\S}[0]{\ensuremath{\hat{S}}}
\newcommand{\Sminuseff}[0]{\ensuremath{\hat{S}_-^{\mathrm{eff}}}}
\newcommand{\Sxeff}[0]{\ensuremath{\hat{S}_x^{\mathrm{eff}}}}
\newcommand{\Syeff}[0]{\ensuremath{\hat{S}_y^{\mathrm{eff}}}}
\newcommand{\tildeac}[0]{\ensuremath{\tilde{a}_{\mathrm{c}}}}
\newcommand{\tildepauli}[0]{\ensuremath{\tilde{\sigma}}}
\newcommand{\Tcaveff}[0]{\ensuremath{T_{\mathrm{cav}}^{\mathrm{eff}}}}
\newcommand{\Techo}[0]{\ensuremath{T_{\mathrm{echo}}}}
\newcommand{\Tmem}[0]{\ensuremath{T_{\mathrm{mem}}}}
\newcommand{\Tswap}[0]{\ensuremath{T_{\mathrm{swap}}}}
\newcommand{\Var}[0]{\ensuremath{\mathrm{Var}}}
\renewcommand{\vec}[1]{\ensuremath{\mathbf{#1}}}
\newcommand{\Xa}[0]{\ensuremath{\hat{X}_{\mathrm{c}}}}

\title{Reaching the quantum limit of sensitivity in electron spin resonance}

\author{A. Bienfait$^{1}$, J.J. Pla$^{2}$, Y. Kubo$^{1}$, M. Stern$^{1,3}$, X. Zhou$^{1,4}$, C.C. Lo$^{2}$, C.D. Weis$^{5}$, T. Schenkel$^{5}$, M.L.W. Thewalt$^{6}$, D. Vion$^{1}$,
D. Esteve$^{1}$, B. Julsgaard$^{7}$, K. Moelmer$^{7}$, J.J.L. Morton$^{2}$, and P. Bertet$^{1}$}

\affiliation{$^{1}$Quantronics group, Service de Physique de l'Etat Condens\'e, DSM/IRAMIS/SPEC, CNRS UMR 3680, CEA-Saclay,
91191 Gif-sur-Yvette cedex, France }

\affiliation{$^{2}$ London Centre for Nanotechnology, University College London, London WC1H 0AH, United Kingdom}
	
\affiliation{$^{3}$ Quantum Nanoelectronics Laboratory, BINA, Bar Ilan University, Ramat Gan, Israel}
	
\affiliation{$^{4}$Institute of Electronics Microelectronics and Nanotechnology, CNRS UMR 8520, ISEN Department, Avenue Poincar\'e, CS 60069, 59652 Villeneuve d'Ascq Cedex, France}

\affiliation{$^{5}$Accelerator Technology and Applied Physics Division, Lawrence Berkeley National Laboratory, Berkeley,
California 94720, USA}

\affiliation{$^{6}$Dept. of Physics, Simon Fraser University, Burnaby, British Columbia V5A 1S6, Canada}

\affiliation{$^{7}$Department of Physics and Astronomy, Aarhus University, Ny Munkegade 120, DK-8000 Aarhus C, Denmark}


\date{\today}

\maketitle

{\bf The detection and characterization of paramagnetic species by electron-spin resonance (ESR) spectroscopy is widely used throughout chemistry, biology, and materials science~\cite{SchweigerEPR(2001)}, from in-vivo imaging~\cite{yoshimura1996vivo} to distance measurements in spin-labeled proteins~\cite{duss2014epr}. ESR typically relies on the inductive detection of microwave signals emitted by the spins into a coupled microwave resonator during their Larmor precession --- however, such signals can be very small, prohibiting the application of ESR at the nanoscale, for example, at the single-cell level or on individual nanoparticles. In this work, using a Josephson parametric microwave amplifier combined with high-quality factor superconducting micro-resonators cooled at millikelvin temperatures, we improve the state-of-the-art sensitivity of inductive ESR detection by nearly $4$ orders of magnitude. We demonstrate the detection of 1700 bismuth donor spins in silicon within a single Hahn~\cite{Hahn.PhysRev.80.580(1950)} echo with unit signal-to-noise (SNR) ratio, reduced to just 150 spins by averaging a single Carr-Purcell-Meiboom-Gill sequence~\cite{Mentink.JMagRes.236.117(2013)}. This unprecedented sensitivity reaches the limit set by quantum fluctuations of the electromagnetic field instead of thermal or technical noise, which constitutes a novel regime for magnetic resonance. The detection volume of our resonator is $\sim$0.02~nl, and our approach can be readily scaled down further to improve sensitivity, providing a new and versatile toolbox for ESR at the nanoscale.}

A wide variety of techniques are being actively explore to push the limits of sensitivity of ESR to the nanoscale, including approaches based on optical~\cite{Wrachtrup.Nature.363.244(1993),Grinolds.NatureNano.9.279(2014)} or electrical~\cite{Hoehne.RevSciInst.4.043907(2012),Morello.Nature.467.687(2010)} detection, as well as scanning probe methods~\cite{Manassen.PhysRevLett.62.2531(1989),Rugar.Nature.360.563(1992)}. Our focus in this work is to maximise the sensitivity of inductively detected pulsed ESR, in order to maintain the broad applicability to different spin species as well as fast high-bandwidth detection. Pulsed ESR spectroscopy proceeds by probing a sample coupled to a microwave resonator of frequency $\omega_0$ and quality factor $Q$ with sequences of microwave pulses that perform successive spin rotations, triggering the emission of a microwave signal called a spin-echo whose amplitude and shape contain the desired information about the number and properties of paramagnetic species. The spectrometer sensitivity is conveniently quantified by the minimal number of spins $N_{min}$ that can be detected within a single echo~\cite{Hahn.PhysRev.80.580(1950)}. Conventional ESR spectrometers use $3$D resonators with moderate quality factors in which the spins are only weakly coupled to the microwave photons and thus obtain a sensitivity of $N_{min} \sim 10^{13}$ spins at $T = 300$\,K and X-band frequencies ($\omega_0 / 2\pi \sim 9-10\,$GHz). To increase the sensitivity, micro-fabricated metallic planar resonators with smaller mode volumes have been used, resulting in larger spin-microwave coupling~\cite{Wallace.RSI.62.1754(1991),Narkowicz.JMagRes.175.275(2005)}. Combined with operation at $T = 4$\,K and the use of low-noise cryogenic amplifiers and superconducting high-Q thin-film resonators, sensitivities up to $N_{min} \sim 10^{7}$ spins have been reported, which represents the current state-of-the-art~\cite{Benningshof201384(2013),Sigillito.APL.104.104.22407(2014),Artzi.APL.106.084104(2015)}. 

Further improvements in the sensitivity of ESR spectroscopy can be obtained by cooling the sample and resonator down to mK temperatures that satisfy $T \ll \hbar \omega_0 / k_B$ at X-band frequencies. As a result, both the spins and the microwave field reach their quantum ground state, which is the optimal situation for magnetic resonance since the spins are then fully polarized and thermal noise suppressed. The noise in the emitted echo signal is essentially due to vacuum quantum fluctuations of the microwave field, with a dimensionless spectral power density $n_{eq} = S(\omega) / (\hbar \omega) = 1 / 2$, possibly supplemented by extra noise $n_s$ due to spontaneous emission of the spins (see Supplementary Information). The total noise spectral density in the detected signal $n = n_{eq} + n_{s} + n_{amp}$ however also includes the added noise $n_{amp}$ of the first amplifier of the detection chain; benefiting from the low noise afforded by low temperature operation thus requires nearly noiseless amplifiers at microwave frequencies, as were recently developed in the context of superconducting quantum circuits. These Josephson Parametric Amplifiers (JPAs) are operated at mK temperatures, have a bandwidth of up to $\approx 100$\,MHz, and a low saturation input power (typically $1-10$\,fW)~\cite{bergeal_phase-preserving_2010,Zhou.PhysRevB.89.214517(2014)}. They have been shown to add the minimum amount of noise permitted by quantum mechanics~\cite{Caves.PhysRevD.26.1817(1982)}: $n_{amp} = 0.5$ when both field quadratures are equally amplified (non-degenerate mode)~\cite{bergeal_phase-preserving_2010}, and $n_{amp} = 0$ when only one quadrature is amplified (degenerate mode)~\cite{Castellanos.NaturePhys.4.929(2008)}. JPAs have been used so far for reading-out the state of superconducting qubits~\cite{Vijay.PhysRevLett.106.110502(2011)}, the motion of nanomechanical oscillators~\cite{Teufel.NatureNano.4.820(2009)} and the charge state of a quantum dot~\cite{Stehlik.arxiv.1403.3871(2014)}, as well as for high-sensitivity magnetometry~\cite{Hatridge.PhysRevB.83.134501(2011)}. Here we show that they are also well suited to amplify the weak and narrow-band signals emitted by small numbers of spins, with the ultimate sensitivity allowed by quantum mechanics, enabling us to demonstrate a $4$ orders of magnitude improvement in sensitivity over the state-of-the-art.

\begin{figure}[!htbp]
  \centering
  \includegraphics[width=160mm]{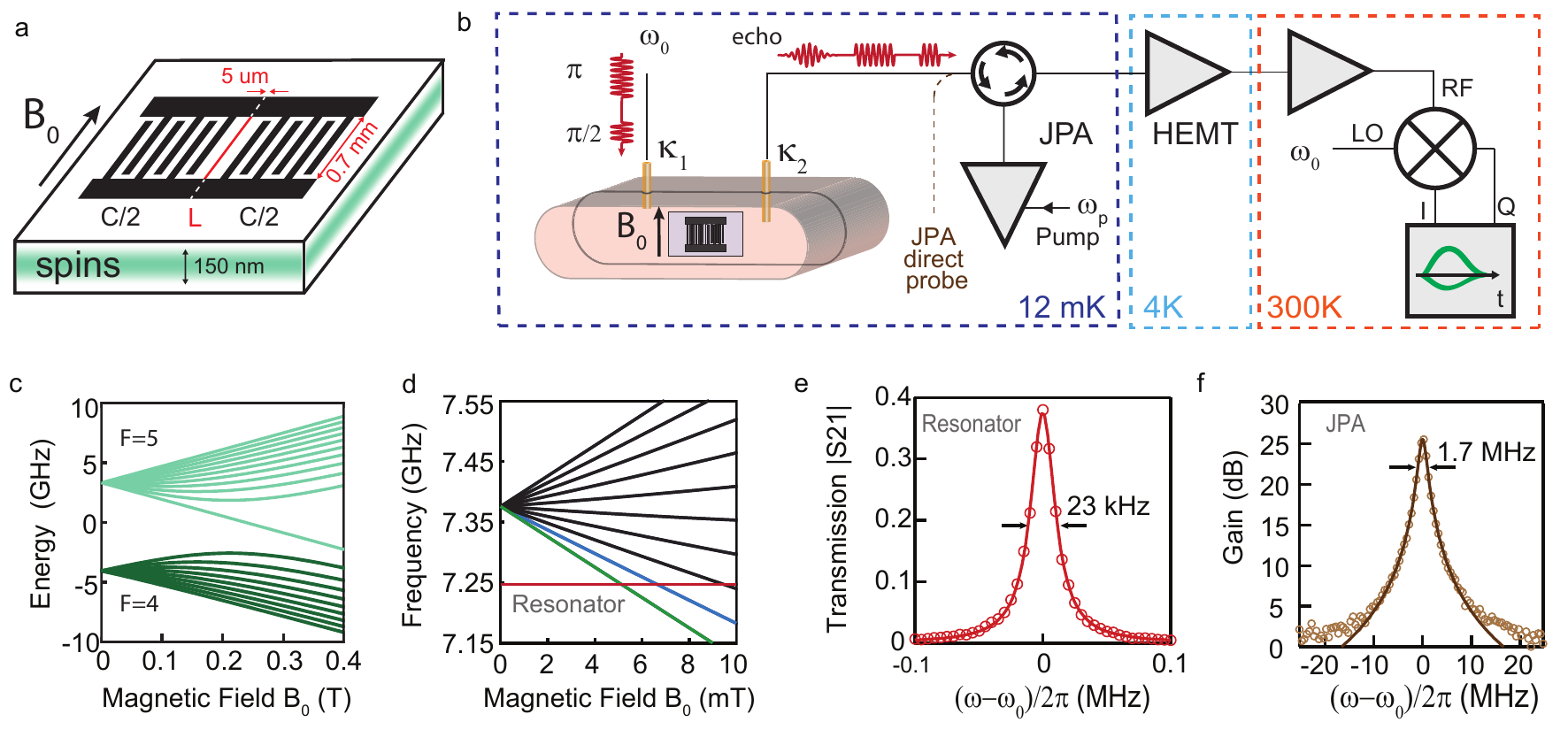}
  \caption{{\bf Experimental setup and spin system.} (a) The aluminium microwave resonator with frequency $\omega_0$ consists of an interdigitated capacitor in parallel with a $5 \mu \mathrm{m}$-wide wire inductor, fabricated on a Bi-doped $^{28}\mathrm{Si}$ epi-layer.
(b) The sample is mounted in a copper box, thermally anchored at $12$\,mK, and probed by microwave pulses via asymmetric antennas coupled  to the resonator with rate $\kappa_1=1.2\cdot 10^4 \mathrm{s}^{-1}$ and $\kappa_2=5.6\cdot 10^4 \mathrm{s}^{-1}$. A magnetic field $B_0$ is applied parallel to the resonator inductance. Microwave pulses at $\omega_0$ are sent by antenna $1$, and the microwave signal leaving via antenna $2$ is directed to the input of a Josephson Parametric Amplifier (JPA). The JPA is powered by a pump signal at $\omega_p \approx 2 \omega_0$, and its output is further amplified at $4$\,K by a High Electron-Mobility Transistor amplifier, followed by amplification and demodulation at room-temperature, yielding the two field quadratures $I(t),Q(t)$. 
(c) Energy levels of Bi donors in Si, expressed in frequency units (see spin Hamiltonian in the Supplementary Material). (d) ESR-allowed transitions in the low-field limit. For $B_0 \leq 8$\,mT, the $\ket{F,m_F} = \ket{4,-4} \rightarrow \ket{5,-5}$ and $\ket{4,-3} \rightarrow \ket{5,-4}$ transitions cross the resonator frequency at respectively $B_0=5$ and $7$\,mT.
(e) Measured resonator transmission coefficient $|S_{21}|$ (red circles), yielding $\omega_0 / 2\pi = 7.24$\,GHz and a total quality factor $Q=3 \cdot 10^5$ (red curve is a fit). 
(f) The JPA can be characterized via a direct line bypassing the resonator, yielding a gain, in non-degenerate mode, of $G>20$\,dB above a $3$\,MHz bandwidth. Circles are experimental data, curve is a Lorentzian fit.}
		\label{fig1}
	\end{figure}

We use an ensemble of bismuth donors implanted over a $150$\,nm depth into an isotopically enriched silicon-28 crystal, on top of which we pattern a superconducting aluminium thin-film micro-resonator consisting of an interdigitated capacitor in parallel with a wire inductance (see Fig.~\ref{fig1} for a sketch of the setup). Due to this geometry, the microwave field $B_1 \cos \omega_0 t$ couples only to the $N_{Bi} \simeq 4 \cdot 10^7$ implanted Bi atoms located in the area below the wire. The sample is inserted inside a copper box to suppress the resonator radiative losses while enabling to probe its transmission by capacitive coupling to input and output antennas. In this well-controlled environment the resonator reaches a loaded quality factor $Q=3 \times 10^5$ for a frequency $\omega_0 / 2\pi = 7.24$\,GHz (see Fig.~\ref{fig1}e). Microwave pulses at $\omega_0$ are applied to the cavity input; the output signal (including the echoes emitted by the spins) is directed towards the input of a JPA~\cite{Zhou.PhysRevB.89.214517(2014)} with power gain $G$ up to $\approx 23$\,dB at $\omega_0$ when powered by a pump microwave signal at a frequency $\omega_p \approx 2 \omega_0$, $\omega_p = 2 \omega_0$ corresponding to the degenerate mode of operation~\cite{Zhou.PhysRevB.89.214517(2014)}. The JPA output is then further amplified by a semiconducting HEMT amplifier at $4$\,K, and finally demodulated at frequency $\omega_0$ yielding time traces for the two quadratures ${I(t),Q(t)}$ (see Supplementary Information for more details). The sample and JPA are cooled at $12$\,mK in a dilution refrigerator.

An in-plane magnetic field $B_0$ is applied parallel to the sample surface along the resonator inductance. Neutral bismuth donors in silicon have a $S=1/2$ electron spin coupled by a strong hyperfine interaction term $A \overrightarrow{S} \cdot \overrightarrow{I}$ to the $I=9/2$ nuclear spin of $^{209}$Bi~\cite{Feher.PhysRev.114.1219(1959),Morley.NatureMat.9.725(2010)}, with $A/h = 1.48$\,GHz. In the low-field regime, the $20$ electro-nuclear energy states are best described by their total angular momentum $\overrightarrow{F}=\overrightarrow{S}+\overrightarrow{I}$ and its projection $m_F$ --- they can be grouped in a $F=4$ ground and a $F=5$ excited multiplet separated by a frequency of $5A/h = 7.38\,$GHz in zero field (see Fig.~\ref{fig1}). With the chosen orientation of $B_0$, the $B_1$ microwave field generated by the resonator is perpendicular to the spin quantization axis and only transitions obeying $|\Delta m_F| =1$ have a significant matrix element (see Fig.~\ref{fig1}d) for $B_0 \leq 10$\,mT. Their frequency in the $\sim 7.3-7.5$\,GHz range makes Bi:Si an ideal system for coupling to superconducting aluminum resonators which can withstand only fields below $\simeq 10$\,mT.

\begin{figure}[!htbp]
  \centering
  \includegraphics[width=85mm]{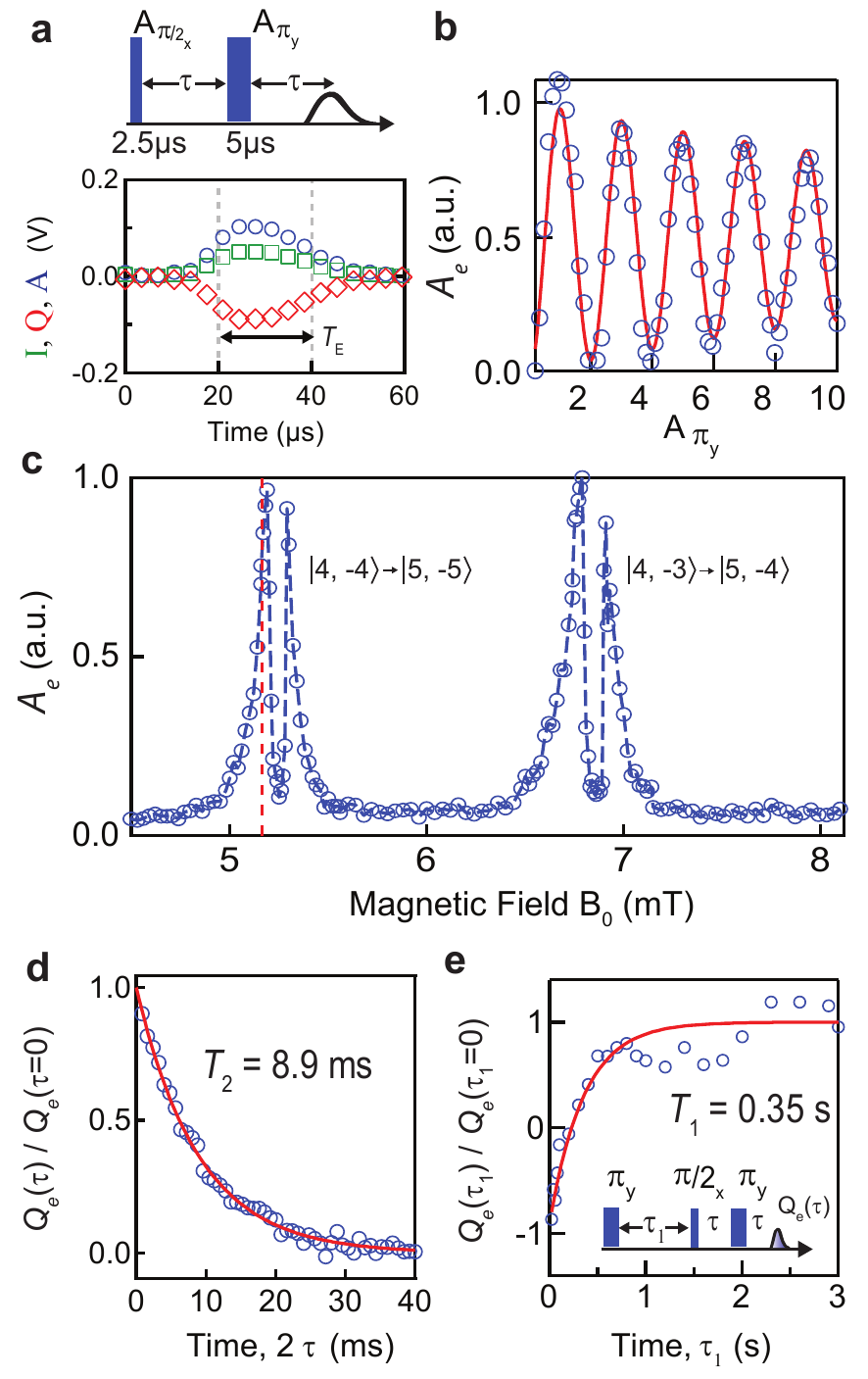}
  \caption{{\bf Sample characterization}. (a) Hahn-echo sequence (top), triggering the emission of an echo (bottom). Plotted are the demodulated quadratures $I(t)$ (green squares) and $Q(t)$ (red diamond), as well as the echo amplitude $A(t)=\sqrt{I(t)^2 + Q(t)^2}$ (blue circles), from which the echo quadrature area $X_e = \int_{-T_E/2}^{+T_E/2} X(t)dt$ (with $X=I,Q$) and amplitude area $A_e = \int_{-T_E/2}^{+T_E/2} A(t)dt$ are extracted. The data were taken for $B_0 = 5.2$~mT.
 (b) Normalised amplitude echo area as a function of the refocusing pulse amplitude $A_\pi$ (rescaled by the amplitude needed for a $\pi$ pulse) showing Rabi oscillations. Blue circles are data points, red curve is an exponentially damped sine fit.
 %
 (c) Amplitude echo area (blue circles joined by dashed lines) as a function of magnetic field $B_0$ showing two principal resonances, each split into a doublet due to the effect of strain on the donors below and next to the aluminium wire inductor. 
 (d) As the total time $2\tau$ between the initial $\pi / 2$ pulse and the echo is increased, the recovered $Q$ quadrature echo area decays with an exponential behaviour (red curve is a fit), yielding a spin coherence time  $T_2 = 8.9$\,ms.
 (e) The inversion recovery sequence (see inset) is used to measure the spin relaxation time $T_1 = 0.37$\,s.
Red curve is an exponential fit to the experimental data (blue circles).}
		\label{fig2}
	\end{figure}

In our case, the $\ket{F,m_F} = \ket{4,-4} \rightarrow \ket{5,-5}$ and $\ket{4,-3} \rightarrow \ket{5,-4}$ transitions are expected to be resonant with $\omega_0$ at $B_0 = 5$ and $7$\,mT respectively; corresponding peaks in the integrated spin-echo signal (of duration $T_E \approx 20 \, \mu \mathrm{s}$) are indeed measured as shown in Fig.~\ref{fig2}a-c. Each transition consists of two sub-peaks, with an inhomogeneous linewidth $\Gamma / 2\pi = 2$\,MHz. We attribute this sub-structure to the differential strain~\cite{Dreher.PhysRevLett.106.037601(2011)} acting on the Bi atoms lying just under the wire versus those around it (see Supplementary Information). We will focus in the following on the $\ket{4,-4} \rightarrow \ket{5,-5}$ transition for the spins lying under the wire, at $B_0 = 5.18$\,mT. Well-defined Rabi oscillations are observed in the integrated echo signal as a function of the refocusing pulse amplitude (see Fig.~\ref{fig2}b), with a $100$\,kHz Rabi frequency for a remarkably low input power of $3$\,pW~\cite{Sigillito.APL.104.104.22407(2014)}. The decay of the integrated echo signal as a function of the total delay $2\tau$ between the initial $\pi / 2$ pulse and the echo is well fitted by an exponential decay with a time constant $T_2=10$\,ms, a typical coherence time for $\mathrm{Bi}:^{28}\mathrm{Si}$~\cite{Wolfowicz.NatureNano.8.561(2013)} (see Fig.~\ref{fig2}d). The energy relaxation time $T_1$ is measured by the inversion recovery method to be $T_1=0.3$\,s (see Fig.~\ref{fig2}e), allowing us to use a $1$\,Hz repetition rate throughout this work.

\begin{figure}[!htbp]
  \centering
  \includegraphics[width=90mm]{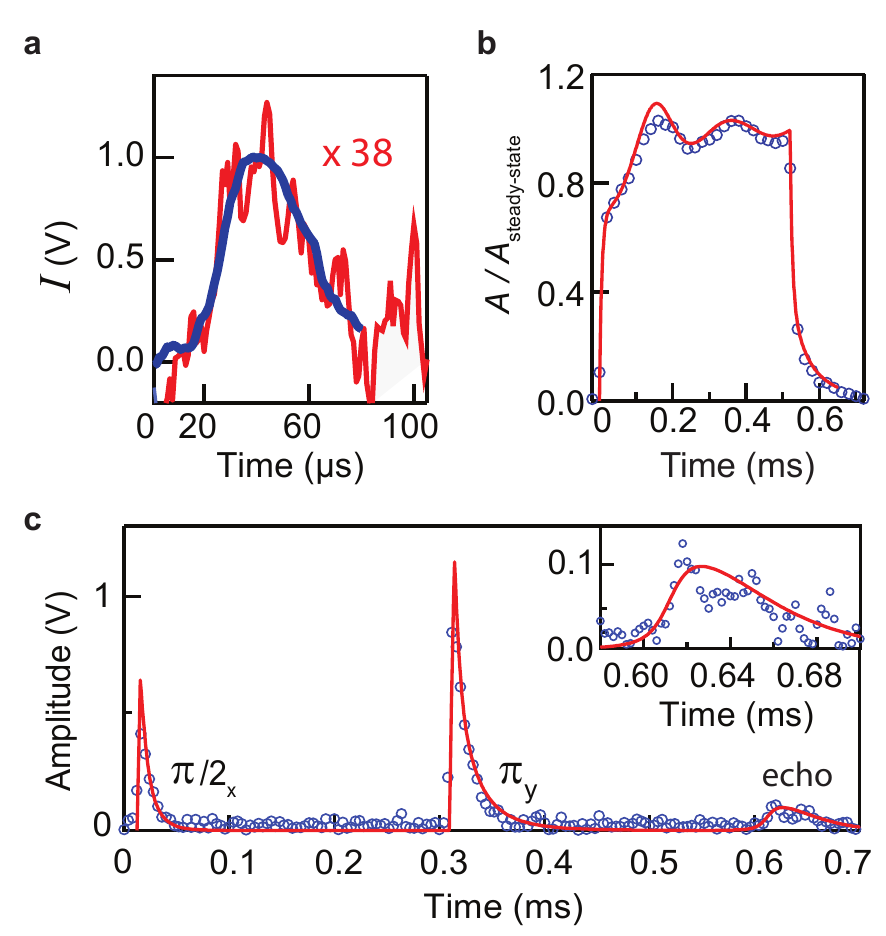}
  \caption{{\bf Spectrometer sensitivity.} (a) Echo signal $I(t)$ without JPA (red curve) and with JPA in degenerate mode and $23$\,dB gain (blue curve), averaged $10$ times. The red curve was rescaled by the amplifier amplitude gain for comparison with the JPA-on curve. The JPA-on (JPA off) echo has a signal-to-noise ratio of $22 \pm 3$ ($2 \pm 0.5$), which translates into a single-echo signal-to-noise of $7 \pm 1$ ($0.6 \pm 0.15$). (b) Time dependent amplitude of a $500 \mu \mathrm{s}$ pulse at $\omega_0$, averaged $1000$ times, showing Rabi oscillations (circles). A simulation (curve) is used to estimate the number of spins contributing to the absorption. (c) Measured microwave amplitude (circles) during an entire Hahn echo sequence, with the JPA  turned off in order to avoid any saturation effect.  A simulation (curve) uses only the number of spins extracted from (b) and shows quantitative agreement with the measurements. These simulations indicates that the $\pi /2$ pulse acts on $1.2 \cdot 10^4$ spins.}
		\label{fig3}
	\end{figure}

The spectrometer sensitivity is estimated by measuring the SNR of a single echo. The JPA is operated in the degenerate mode, with the phase of the pump signal chosen such that the echo signal is entirely on the amplified quadrature. With these optimal settings, the amplitude SNR of the echo shown in Fig.~\ref{fig3}a is found to be $7 \pm 1$, one order of magnitude larger than the SNR obtained in the same conditions but with the JPA pump turned off so that it simply reflects the echo signal. This improvement is consistent with a noise reduction from $n \sim 50$ (with JPA off) down to $n \sim 0.5$, thus close to the quantum limit, and with calibration measurements performed on the JPA itself (see Supplementary Information).

\begin{figure}[!htbp]
  \centering
  \includegraphics[width=89mm]{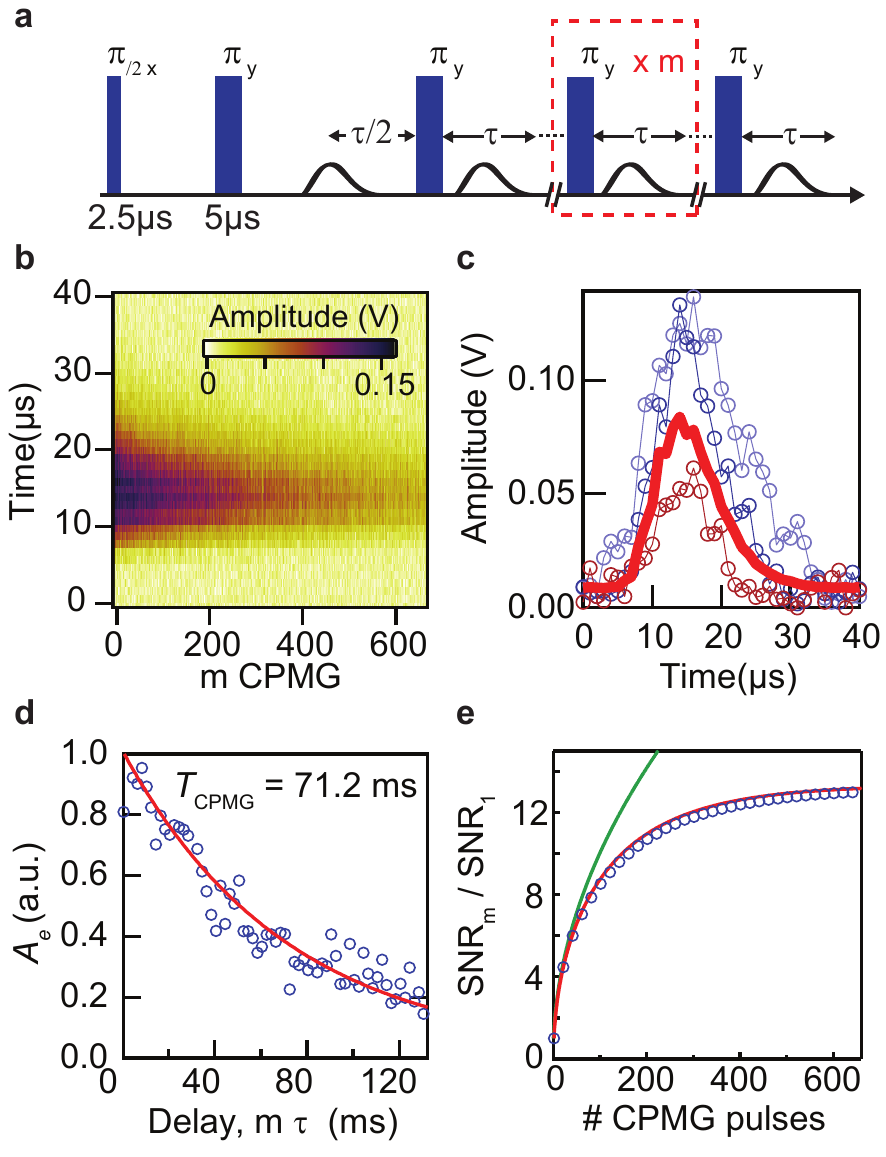}
  \caption{{\bf Further sensitivity improvement with the Carr-Purcell-Meiboom-Gill (CPMG) pulse sequence}. (a) A spin echo generated by any pulsed ESR experiment can be refocused by 
a train of $\pi$ pulses with rotations axes oriented along the echo phase direction, and thus used to enhance the SNR in a single shot. 
(b) Time-dependence of the measured echo amplitude as a function of the echo number, $m$, over the course of a single sequence comprising $650$ $\pi$ pulses separated by $\tau = 200 \mu \mathrm{s}$. 
%
(c) Three time traces (circles) of echoes number 1, 100, 600 are explicitly shown from (b). Solid line shows the average over all 650 echoes.
(d) Decay of the echo area $A_e$  as a function of the total delay between the initial $\pi / 2$ pulse and the echo (circles), fitted with an exponential decay (curve) of time constant $T_{\rm CPMG} = 71$\,ms. 
(e) SNR improvement as a function of the number $m$ of $\pi$ pulses within the CPMG sequence (circles), showing a tenfold improvement. In the absence of decoherence, the SNR should follow $\sqrt{m}$ (green curve); with decoherence (red curve) the SNR levels off, and eventually decays for higher $m$.}
		\label{fig4}
	\end{figure}

Of all the neutral Bi donors within the resonator mode volume, only those whose frequency lies within the resonator linewidth $\kappa = \omega_0 / Q$ and which are in the $\ket{4,-4}$ state contribute to the echo signal. A rough estimate of the number of spins is therefore obtained as $N_{Bi} (\kappa / \Gamma) / 9 = 4 \cdot 10^4$, an overestimate given that only a fraction of implanted atoms shows a magnetic resonance signal due to either crystal damage or to donor ionization~\cite{Weis.APL.100.172104(2012)}. For a more accurate determination, the time-dependent absorption of a microwave pulse at $\omega_0$ recorded and fitted to a simple model (see Fig.~\ref{fig3}b and Supplementary Information) allows us to obtain an abolute calibration of the spin density. A whole spin-echo sequence is then measured and simulated (see Fig.~\ref{fig3}c); the quantitative agreement with the observed echo amplitude establishes (from the simulations) that $1.2 \times 10^4$ spins are excited during the sequence. This implies a $\sim 30 \%$ yield between number of implanted atoms and of neutral donors, compatible with previous reports~\cite{Weis.APL.100.172104(2012)}.

Overall, the spectrometer can therefore detect down to $N_{min} = 1.2 \times 10^4 / 7 = 1.7 \cdot 10^3$ spins with a signal-to-noise of unity in a single Hahn-echo, and has a corresponding sensitivity of $1.7 \cdot 10^3 \,\mathrm{spins}/\sqrt{\mathrm{Hz}}$ given the $1$\,Hz repetition rate. This $4$ orders of magnitude improvement over the state-of-the-art is in qualitative agreement with the prediction of a simplified model (see Supplementary Information) $N_{min}^{(th)} \simeq \sqrt{\frac{n \kappa}{T_E}} \frac{1}{g}$, $g$ being the coupling constant of a single spin to the resonator microwave field, estimated for our geometry to be $g / 2\pi = 55$\,Hz, which yields $N_{min}^{(th)} = 400$ spins. The sensitivity can be further improved with a Carr-Purcell-Meiboom-Gill pulse sequence, adding $m$ $\pi_Y$ pulses after the first echo in order to recover $m$ echoes instead of a single one, yielding an increase in signal-to-noise of $\approx \sqrt{m}$~\cite{Mentink.JMagRes.236.117(2013)}. The applicability of this technique depends on factors such as the spin coherence time $T_2$ of the sample and the echo duration $T_E$ --- for our $^{28}$Si:Bi sample up to $600$ echoes are obtained, as shown in Fig.~\ref{fig4}, with a corresponding tenfold increase of the SNR and an unprecedented sensitivity of $150 \,\mathrm{spins}$ in a single shot or $150 \,\mathrm{spins} /\sqrt{\mathrm{Hz}}$. 

A wide range of species including molecular magnets, Gd spin-labels and high-spin defects in solids, can be studied by ESR at low magnetic fields using the Al thin-film resonator demonstrated here. Operation in larger magnetic fields ($\sim 0.3$\,T) would enable the most general application of this method to other spin species and could be achieved by fabricating the micro-resonator from higher critical field superconductors such as Nb~\cite{Sigillito.APL.104.104.22407(2014)} or NbTiN~\cite{Ranjan.PhysRevLett.110.067004}. Our results thus open the way to performing ESR spectroscopy on nanoscale samples such as single cells, small molecular ensembles, nanoparticles and nano-devices. We predict a further $2$ orders of magnitude sensitivity enhancement is possible by reducing the resonator transverse dimensions down to the nanometric scale, which would then be sufficient for detecting individual electron spins.

\textbf{Acknowledgements} We acknowledge technical support from P. S{\'e}nat, D. Duet, J.-C. Tack, P. Pari, P. Forget, as well as useful discussions within the Quantronics group. We acknowledge support of the European Research Council under the European Community's Seventh Framework Programme (FP7/2007-2013) through grant agreements No. 615767 (CIRQUSS), 279781 (ASCENT), and 630070 (quRAM), and of the C'Nano IdF project QUANTROCRYO. J.J.L.M. is supported by the Royal Society. C.C. Lo is supported by the Royal Commission for the Exhibition of 1851. B. Julsgaard and  K. M{\o}lmer acknowledge support from the Villum Foundation. 

\hrulefill

\newpage
\widetext
\begin{center}
\textbf{\large Supplementary Material: Reaching the quantum limit of sensitivity in electron spin resonance}
\end{center}
\setcounter{equation}{0}
\setcounter{figure}{0}
\setcounter{table}{0}
\setcounter{page}{1}
\makeatletter
\renewcommand{\theequation}{S\arabic{equation}}
\renewcommand{\thefigure}{S\arabic{figure}}

%

\newcommand{\ain}[0]{\ensuremath{\hat{a}_{\mathrm{in}}}}
\newcommand{\aout}[0]{\ensuremath{\hat{a}_{\mathrm{out}}}}
\newcommand{\bid}[0]{\ensuremath{\hat{b}_{\mathrm{id}}}}
\renewcommand{\c}[0]{\ensuremath{\hat{c}}}
\renewcommand{\H}[0]{\ensuremath{\hat{H}}}
\renewcommand{\Im}[0]{\ensuremath{\mathrm{Im}}}
\newcommand{\kappaL}[0]{\ensuremath{\kappa_{\mathrm{L}}}}
\newcommand{\kB}[0]{\ensuremath{k_{\mathrm{B}}}}
\newcommand{\namp}[0]{\ensuremath{n_{\mathrm{amp}}}}
\renewcommand{\neq}[0]{\ensuremath{n_{\mathrm{eq}}}}
\newcommand{\Nmin}[0]{\ensuremath{N_{\mathrm{min}}}}
\newcommand{\nsp}[0]{\ensuremath{n_{\mathrm{sp}}}}
\renewcommand{\Re}[0]{\ensuremath{\mathrm{Re}}}
\renewcommand{\S}[0]{\ensuremath{\hat{S}}}
\renewcommand{\vec}[1]{\ensuremath{\mathbf{#1}}}
\newcommand{\Xid}[0]{\ensuremath{\hat{X}_{\mathrm{id}}}}
\newcommand{\Xin}[0]{\ensuremath{\hat{X}_{\mathrm{in}}}}
\newcommand{\Xout}[0]{\ensuremath{\hat{X}_{\mathrm{out}}}}
\newcommand{\Yin}[0]{\ensuremath{\hat{Y}_{\mathrm{in}}}}
\newcommand{\Yout}[0]{\ensuremath{\hat{Y}_{\mathrm{out}}}}

\section{Experimental details}
\subsection*{Bismuth implanted sample}

\begin{figure}[!htbp]
  \centering
  \includegraphics[scale=2]{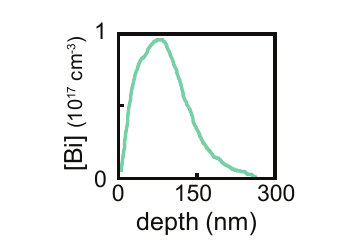}
  \caption{Bismuth implantation profile, measured via Secondary Ion Mass Spectroscopy (SIMS)}
		\label{figBiSiConc}
\end{figure}

The sample consists of a natural silicon (100) substrate on which a $700\,$nm-thick isotopically enriched $99.95\%$ $^{28}$Si epitaxial layer was grown. Bismuth dopants were subsequently implanted into the epitaxial layer and activated by thermal annealing (see~\cite{Weis.APL.100.172104(2012)} for more details). The implantation profile, measured via Secondary Ion Mass Spectroscopy (SIMS) is shown in \ref{figBiSiConc} of the main text. The activation step consists in an anneal to $800 ^{\circ} C$ for $20$\,min under nitrogen atmosphere. An electrical activation yield of 60\% has been measured using a Hall effect measurement system under similar implantation conditions~\cite{Weis.APL.100.172104(2012)}.

A 50-nm-thick aluminum resonator was deposited on top of the sample using a standard lift-off process. The resonator is a lumped element constituted by a $5\mu \mathrm{m}$-wide wire and an interdigitated capacitance of $12$ $50\mu \mathrm{m}$-wide fingers spaced by $50\,\mu \mathrm{m}$. The wire is $730\,\mu \mathrm{m}$ long yielding a resonance $\omega_0/ 2\pi = 7.26\,$GHz.

\subsection*{Measurement setup}

The detailed microwave setup is shown in supplementary Fig.~\ref{fig1Supp}. The sample is enclosed in a box made of oxygen-free-high-conductivity-copper, whose lowest resonance mode is at $8.4\,$GHz. Its role is to suppress the resonator radiative losses, while enabling its coupling to the input and output antennas with rates $\kappa_1$ (input) and $\kappa_2$ (output). The values of $\kappa_1$ and $\kappa_2$ as well as the resonator internal losses $\kappa_L$ are determined experimentally by measuring the complete resonator scattering matrix with the spins far from resonance, and fitting this matrix to the known resonator input-output formulas (see~\cite{PalaciosLaloy2010} for instance). The results are shown in Table~\ref{table1}. As can be seen in Table~\ref{table1}, the experiment is in the so-called critical coupling regime where the internal losses $\kappa_L$ are approximately equal to the external coupling $\kappa_1+\kappa_2$. The asymmetry between $\kappa_1$ and $ \kappa_2$ was purposedly chosen to be large ($\approx 5$) so as to ensure that the majority of the photons emitted by the spins would be collected by the output antenna. The resonator total quality factor is $Q=\dfrac{\omega_0}{\kappa_1+\kappa_2+\kappa_L}=3 \times 10^5$. 

\begin{table}[!htbp]
\setlength{\tabcolsep}{15pt}
\centering
\begin{tabular}{c|c|c|c|c}

$\frac{\omega_r}{2\pi}$ & Q & $\kappa_1$ & $\kappa_2$ & $\kappa_L$  \\ 
\hline 
$7.24\,$GHz & $3 \times 10^5$ & $2.1\times10^3\,$s$^{-1}$ & $9.2\times10^3\,$s$^{-1}$ & $12\times10^3\,$s$^{-1}$  \\ 

\end{tabular} 
  \caption{}
		\label{table1}
\end{table}

During a spin-echo sequence, a series of microwave pulses at frequency $\omega_0$ is sent on the input line of the cavity ($\kappa_1$, green wire on the figure). The transmitted signal is then routed to the Josephson Parametric Amplifier (JPA) via a circulator and is further amplified first by a low-noise HEMT amplifier at the 4K stage then at room temperature, before being finally demodulated by mixing with the local oscillator at frequency $\omega_0$.

\begin{figure}[!htbp]
  \centering
  \includegraphics[width=180mm]{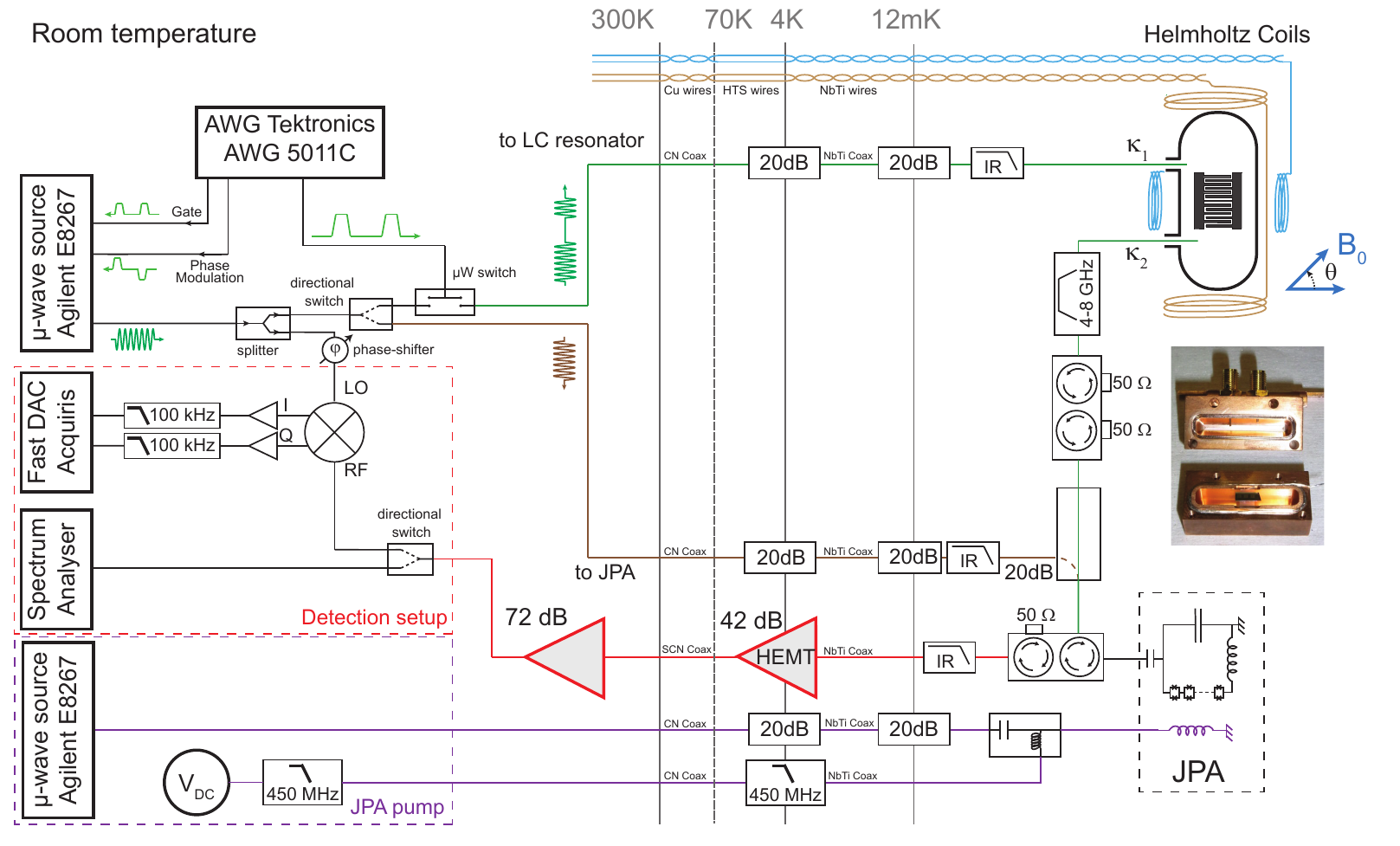}
  \caption{Measurement setup.}
		\label{fig1Supp}
\end{figure}

The JPA can be studied and tuned independently from the cavity by an additional input line (brown) coupled via a 20dB coupler to the cavity output line. Its design and operation have been described in detail in~\cite{Zhou.PhysRevB.89.214517(2014)}. It consists of a lumped element resonator formed by an interdigitated capacitance, a geometrical inductance and an array series of $8$ superconducting  quantum interference devices (SQUIDs). The SQUID array acts as a flux-tunable inductor that allows the resonator frequency to be tuned over a $400\,$MHz frequency range by passing a dc current through an on-chip antenna. The amplifier is parametrically pumped by modulating the flux threading the SQUIDs at a frequency $\omega_p$ close to twice the resonator frequency. The JPA can be operated either in phase sensitive mode where $\omega_p/2=\omega_0$ or in non-degenerate mode. In the latter mode, around $7.3\,$GHz, a non-degenerate gain of $23\,$dB can be obtained with the appropriate pump power as shown in Fig.~1 of the main text. Saturation of the JPA occurs for a typical input power of $-130\,$dBm. In all the data presented in this work, the echo signals emitted by the spins are below the amplifier saturation threshold. The detuning between the pump and the signal is chosen to be $\approx 500\,$kHz, and the demolutated signal is filtered at $100\,$kHz to suppress the idler. In phase-sensitive mode, only one signal quadrature is amplified, and the JPA has an additional $6$\,dB gain. The quadrature is chosen by tuning the relative phase of the pump and signal sources. In addition, the JPA pump is pulsed via the microwave source internal switch (not shown on schematic= so as to generate gain only during the emission of an echo signal. This is done to reduce the effective pump power brought to the mixing chamber plate and thus bring the cryostat temperature from $20$\,mK with a continuous pump signal to $12$\,mK.

A double circulator is used to prevent interferences between the cavity and the JPA. Another double circulator is needed at the JPA output to ensure both the routing of the signal and the isolation from the thermal photons travelling down the output line (red line). Leakage of the $14.5$\,GHz pump to the resonator and the spins is suppressed by a $4-8\,$GHz bandpass filter inserted between the cavity and the JPA. Each input line is attenuated by 20dB at 4K and 20dB at 12mK to thermalize the electromagnetic field, and filtered by low-pass filters containing infra-red absorptive material in order to minimize losses due to out-of-equilibrium quasi-particles generated in the superconducting thin-film. Both the cavity and the JPA are magnetically shielded. The JPA is enclosed in an 3-cm-wide aluminium box surrounded by a 1-mm-thick cryoperm material, the whole being placed inside a 20-cm-long $\mu$-metal cylindrical shield. The magnetic field $B_0$ is applied parallel to the sample surface with an arbitrary angle $\theta$ with respect to the resonator axis, by using two orthogonal Helmholtz coil pairs that can provide up to $10$\,mT and have been calibrated in a previous experiment.

Pulses are shaped by a microwave switch in series with the microwave source internal gate. The relative phases of the pulses are controlled by analog phase modulation. Every control signal is generated by an arbitrary wavefrom generator (AWG). In order to suppress any offset in the detection chain without time-consuming calibration, every pulse sequence is repeated twice with opposite phases on the $\pi /2$ pulses. This phase cycling protocol yields two echo signals with opposite phases taken in the same conditions : the offset is removed by taking the difference between the two time traces acquired on each sequence.

The spin energy relaxation time $T_1$ being $\sim 0.4\,$s (see Fig.2e of the main text), we choose a repetition rate $\gamma_{rep}$ sufficiently slow to allow full relaxation of the spins in-between successive sequences. For example, the spin-echo spectroscopy shown on Figure 2 is acquired with $\gamma_{rep}=0.04\,$Hz, the data of Figure 3b\&c with $\gamma_{rep}=0.1\,$Hz and the absorption data of Figure 3d with $\gamma_{rep}=0.3\,$Hz.

\section{Bismuth donor spin and coupling to the resonator}

Neutral bismuth donors in silicon have a $S=1/2$ electron spin and a nuclear spin $I=9/2$ that are strongly coupled by an isotropic hyperfine interaction term $A/2\pi= 1.45\,$ GHz. The system is described with the following Hamiltonian, \cite{Wolfowicz.NatureNano.8.561(2013)}, where $\gamma_e/ 2\pi=28$ GHz/T and $\gamma_n/ 2\pi=7$ MHz/T :

\begin{equation}
  \H /\hbar = \vec{B} \cdot (\gamma_e \vec{S} \otimes \mathbb{1} - \gamma_n \mathbb{1} \otimes \vec{I}) + A\, \vec{S} \cdot \vec{I} 
\label{eq:Hamiltonian_BiSI}
\end{equation}

In the limit of low static magnetic field $B_0$, the 20 electro-nuclear energy states are well approximated by eigenstates of the total angular momentum $\vec{F}=\vec{S}+\vec{I}$,  which can be grouped in an $F=4$ ground and an $F=5$ excited multiplet separated by a frequency of $5A/2\pi=7.35\,$GHz in zero-field, shown on Figure 1. For a given low static field $B_0$ oriented along $\vec{z}$, only transitions verifying $|\Delta m_F|=1$ have a sizeable $S_x $ matrix element (equals to the $S_y$ matrix element) and so may be probed with an excitation field orientated along $\vec{x}$ (or equivalently along $\vec{y}$). We give in the table below details on the two transitions that are accessible to our resonator.

\begin{table}[!h]
\setlength{\tabcolsep}{5pt}
\centering
\begin{tabular}{c|c|c|c}
Transition & Expected crossing field & $df/dB$ & $|\left\langle m_F\mid \hat{S}_x \mid  m'_F \right\rangle | = |\left\langle m_F\mid \hat{S}_y \mid  m'_F \right\rangle |$ \\ 
\hline 
$m_F=-4 \rightarrow -5$ & $5.16\,$mT & $-25.1\,$GHz/T & $0.47$ \\
$m_F=-3 \rightarrow-4$ & $6.68\,$mT & $-19.2\,$GHz/T & $0.42$ \\
\end{tabular} 
  \caption{}
		\label{tbl:Transitions}
\end{table}

\subsection*{Single-spin coupling to the resonator}

\begin{figure}[!htbp]
  \centering
  \includegraphics[width=160mm]{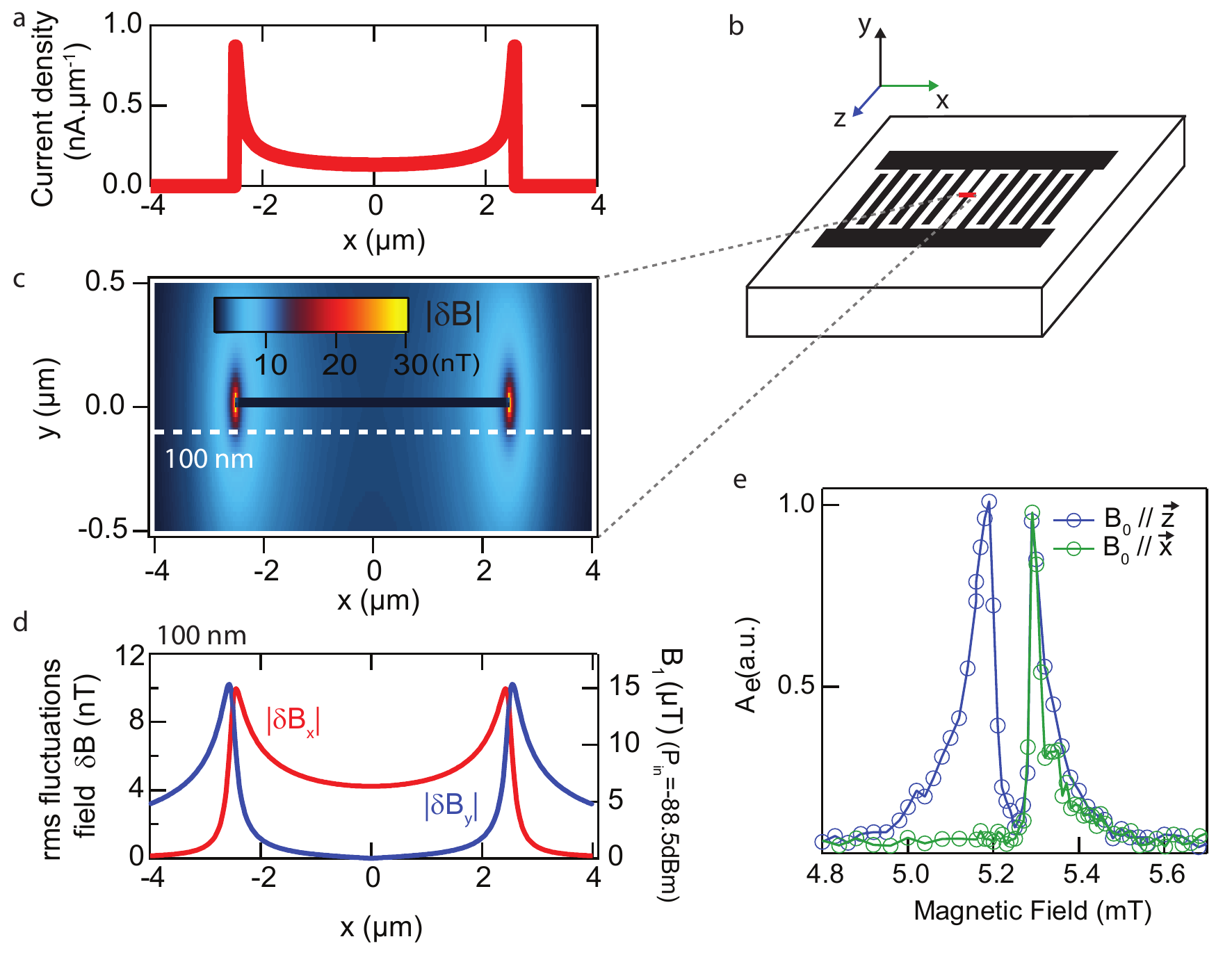}
  \caption{(a) Spatial distribution of the current rms vacuum fluctuations flowing through the resonator inductance, corresponding to an impedance of $44\,\ohm$. (b) Scheme of the resonator, with corresponding directions used in the text. (c)rms vacuum fluctuations of the magnetic field at the given red cross-section on b (d) x (red) and y (blue) components for the vacuum fluctuations of the magnetic field at $y=-100\,$nm (left axis) and for the microwave field  $B_1$ corresponding to an input power $P_{in}=-88.5\,\text{dBm}=1.4\,$pW (right axis). (e) Spin-echo spectroscopy realized for $\mathbf{B_0} = B_0 \cdot \mathbf{z}$ (blue circles)  and $\mathbf{B_0} = B_0 \cdot \mathbf{x}$  (green circles) allowing to make the distinction between spins lying next (strong $\delta B_y$)  and under (strong $\delta B_x$) the aluminium wire.}
		\label{fig2Supp}
\end{figure}

In the experiment, the static magnetic field $B_0$ is applied parallel to the surface (see Fig.~\ref{fig1Supp}) along an axis $\vec{Z}$ that can be decomposed along the orientations defined in Fig.~\ref{fig2Supp} as :

\begin{equation}
  \vec{B_0}=B_0 \vec{Z} = B_0 \cos(\theta) \vec{z} + B_0 \sin(\theta) \vec{x} .
\label{eq:b0_theta}
\end{equation}

A full orthonormal basis is provided by the combination of ${\vec{X},\vec{Y},\vec{Z}}$, with $\vec{X} = \cos \theta \,\vec{x} - \sin \theta\, \vec{z} $, and $\vec{Y} = \vec{y}$. The total magnetic field $\vec{B}$ is the sum of the static bias magnetic field $\vec{B_0}$ and of the microwave field generated by the resonator $\vec{B_1} = \vec {\delta B} (\ac + \adagc)$, where we introduce the magnetic field rms fluctuations at the spin location $\vec {\delta B}$ and the resonator annihilation (resp. creation) operator $\ac$ (resp. $\adagc$). The field $\vec{B_1}$ is located in the plane perpendicular to the resonator wire; as a result one can write $\vec {\delta B} = \delta B_x \vec{x} + \delta B_y \vec{y} = \delta B_X \vec{X} + \delta B_Y \vec{Y} + \delta B_Z \vec{Z}$, with $\delta B_X = \cos \theta \delta B_x$, $\delta B_Y =\delta B_y$ and $\delta B_Z = \sin \theta \delta B_x$. 

Projecting the total Hamiltonian of a single Bismuth donor in the field $\vec{B}$ on the Hilbert space spanned by the two levels ${\ket{m_F}, \ket{m_F'}}$ and introducing the usual Pauli operators yields

\begin{multline}
H / \hbar = \omega_0 \adagc\ac - \frac{\omega_s}{2} \pauli_z \\
 + \gamma_e \left[\langle m_F| S_X \delta B_X + S_Y \delta B_Y | m_F'\rangle \pauli_+ + \langle m_F'| S_X \delta B_X + S_Y \delta B_Y | m_F\rangle \pauli_- \right] (\ac + \adagc) .
\end{multline}

(In this equation we left aside the $S_Z$ terms as they contribute only as negligible fast-rotating terms). As noted in the previous paragraph, $\langle m_F | S_X \delta B_X + S_Y \delta B_Y | m_F' \rangle = \langle m_F| S_X | m_F'\rangle (\delta B_X + i \delta B_Y)$, where $\langle m_F| S_X | m_F'\rangle$ values are shown in Table \ref{tbl:Transitions} ($\approx 0.5$). After performing the rotating-wave approximation, the full system Hamiltonian takes the Jaynes-Cummings form

\begin{equation}
H / \hbar = \omega_0 \adagc \ac - \frac{\omega_s}{2} \pauli_z + g ( \mathrm{e} ^{i \phi_0} \pauli_+ \ac + \mathrm{e} ^{- i \phi_0} \pauli_- \adagc),
\end{equation}

where $\phi_0$ is an irrelevant phase that can be absorbed by a re-definition of the energy levels, and $g$ is the spin-resonator coupling constant given by

\begin{equation}
g = \langle m_F| S_X | m_F'\rangle \gamma_e \sqrt{\delta B_y^2 + (\cos \theta)^2 \delta B_x^2}.
\label{eq:g}
\end{equation}

The vacuum field fluctuations $\vec {\delta B}$ have a spatial dependence, fixed by the shape of the LC resonator mode, which implies that the coupling constant to the resonator will also follow the same spatial dependence. We determine this spatial dependence numerically in the following way. First, the spatial distribution of the current fluctuations in the resonator wire is computed, knowing that the integrated current over the wire cross-sectional area is given by $\delta i = \omega_0 \sqrt{\hbar/2 Z_0}$, $Z_0 = \sqrt{L/C}$ being the resonator impedance estimated to be $44 \Omega$ using the  electromagnetic simulator CST Microwave Studio. For our $50$\,nm-thick aluminum films, the current density is assumed constant in the $y$ direction with an x-dependent integrated value $\delta J(x)$ given by~\cite{VanDuzerTurner(1999)}:

\begin{equation}
\delta J(x) = \begin{cases} \delta J(0)[1-(2x/w)^2]^{-1/2} & \quad \text{for } \lvert x \rvert \leq \lvert\frac{1}{2}w-\lambda^2/(2b) \rvert \\
\delta J(\frac{1}{2}w)\exp{-\left[\left(\frac{1}{2}w-\lvert x \rvert\right)b/\lambda^2\right]} & \quad \text{for } \lvert\frac{1}{2}w-\lambda^2/(2b) \rvert < \lvert x \rvert < \frac{1}{2}w \\
(1.165/\lambda)(wb)^{1/2} \delta J(0) & \quad \text{for } x = \frac{1}{2}w.
\end{cases}
\end{equation}

In these expressions, $w = 5~\mu$m is the width of the wire, $b = 50$~nm is its thickness and $\lambda = 90$~nm is the penetration depth for our Al film. The normalization constant $\delta J(0)$ is determined by the condition that $\int_{-w/2}^{w/2} \delta J(x) dx = \delta i$. From the current distribution, the spatial dependence of $\vec{\delta B}$ is readily obtained using Comsol, and is shown in Fig.\ref{fig2Supp}. Importantly, we note that the $B_1$ field is essentially along  $\vec{x}$ in the region just below the wire, and essentially along  $\vec{y}$ in the region immediately outside of the wire. According to Eq.(5), one thus expects the coupling to the microwave field to be strongly $\theta$-dependent for donors located immediately below the wire, and to depend negligibly on $\theta$ for those outside of the wire.

\subsection*{Sub-structures of the lines profiles}

The spectroscopy of the transition $m_F=-5 \rightarrow m_F=-4$ was realized with the magnetic field $B_0$ applied successively along two directions : along  $\vec{z}$ ($\theta=0\degree$, parallel to the wire), and along $\vec{x}$ ($\theta=90\degree$, perpendicular to the wire). As shown in Supplementary Figure \ref{fig2Supp}, the low-field sub-peak appears only for $\theta=0\degree$ whereas the upper-field structure remains unchanged for both orientations.

The profile of the microwave excitation field components  $B_{1 x}$ and  $B_{1 y}$ at a depth of $100\,$nm (implantation profile peak) is shown in Fig.~\ref{fig2Supp}d. As mentioned earlier the field below the wire is essentially along $\vec{x}$, whereas it is essentially along $\vec{y}$ outside the wire. When $\vec{B_0}=B_0 \vec{z}$, $B_1$ is transverse to $B_0$ for all spins. When $\vec{B_0}=B_0 \vec{x}$, $B_1$ is transverse to $B_0$ only for spins outside the wire. This strongly suggests that when $\vec{B_0}=B_0 \vec{z}$ both spin families can contribute to the signal and when $\vec{B_0}=B_0 \vec{x}$ only spins outside the wire contribute.

\begin{figure}[!htbp]
  \centering
  \includegraphics[scale=1]{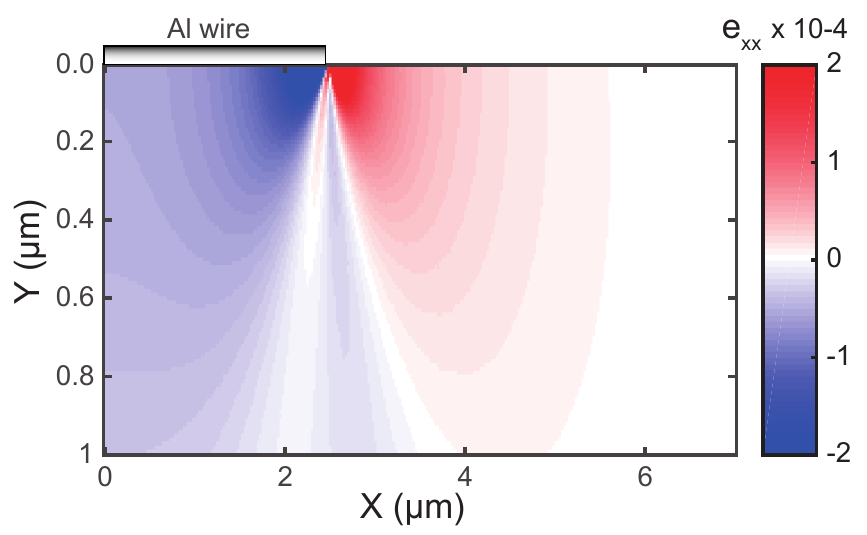}
  \caption{}
		\label{FigSuppStrain}
\end{figure}

As a result of the larger thermal expansion coefficient for aluminium than for silicon, once cooled the region underneath the wire experiences a different strain to the region immediately outside of the wire see Supplementary Figure \ref{FigSuppStrain}. Silicon is an indirect bandgap semiconductor with a six-fold degenerate conduction band minimum. The sharp confining potential of a donor causes the six-fold degenerate ground state to split into a singlet ground state: $A_1 = \frac{1}{\sqrt{6}}\lbrace 1,1,1,1,1,1 \rbrace$,
 a triply-degenerate excited state: $T_{\rm 2(x,y,z)} = \frac{1}{\sqrt{2}} \lbrace 1,-1,0,0,0,0 \rbrace , \frac{1}{\sqrt{2}} \lbrace 0,0,1,-1,0,0 \rbrace , \frac{1}{\sqrt{2}} \lbrace 0,0,0,0,1,-1 \rbrace$ and a doubly-degenerate excited state:
   $E_{\rm (xy,xyz)} = \frac{1}{2} \lbrace 1,1,-1,-1,0,0 \rbrace , \frac{1}{\sqrt{12}} \lbrace -1,-1,-1,-1,2,2 \rbrace$. Strain has the effect of lowering the energy of conduction band minima (or ``valleys'') in the direction of compressive strain and raising the energy of valleys in the direction of tensile strain. Each state will therefore have a shift in energy that depends on the applied strain and its valley composition~\cite{UsmanArxiv(2015)}. In addition, strain mixes the ground state $A_1$ with the doublet excited states $E_{\rm xy}$ and $E_{\rm xyz}$.

The $I = 9/2$ nuclear spin of bismuth means that it possesses an electric quadrupole moment $Q$. The quadrupole moment can interact with an electric field gradient (EFG), which, for example, can be generated by the electron wavefunction through the operator~\cite{Slichter(1990)}:

\begin{align}
V_{\rm \alpha \beta} & = \langle\Psi\vert \widehat{H}^{\rm EFG}_{\rm \alpha \beta} \vert\Psi\rangle \\
\widehat{H}_{\rm \alpha \beta}^{\rm EFG} & = \frac{e}{4\pi\epsilon}\frac{3\alpha\beta-r^2}{r^5}
\end{align}

where $\alpha \text{ or } \beta = x, y, z$ are the crystal or principal coordinate system. In the absence of strain, the ground state electron wavefunction is the $A_1$ state, which is a symmetric combination of the six valleys. This symmetric state produces no EFG and thus a vanishing quadrupole interaction (QI). On the other hand, the excited states are an asymmetric combination of valleys and result in an asymmetric charge distribution and a non-zero EFG. For the case of strain applied along the z principle axis, an EFG is produced through mixing with the doublet excited state $E_{\rm xyz}$. The quadrupole coupling for a field $B_0$ applied in the direction of this EFG is given by the interaction Hamiltonian:
\begin{equation}
\widehat{H}_{\rm QI} = eQV_{\rm z z} \left(3 \widehat{I}_{\rm z}^2 - \overrightarrow{I}(\overrightarrow{I}+1)\right)/\left(4I(2I-1)\right)\label{QI}
\end{equation}

From Equation \eqref{QI}, it is evident that the QI produces an energy shift only on transitions whose states have differing $m_{\rm I}$. This is applicable to our low magnetic field transitions, which are highly mixed in the electron-nuclear basis. The sign of the EFG -- and consequently the quadrupole shift of the spin transitions -- depends on the sign of the induced strain, which as shown in Figure \ref{FigSuppStrain} is opposite for donors underneath the wire and to its side. 

The measured spin resonance lines in our device are split above and below the theoretical field values (see Figure 2 of main manuscript), suggesting an underlying interaction with both positive and negative frequency components. As we have shown, such a frequency distribution could be explained by a strain-induced QI. This is completely consistent with the observation above that the two sub-peaks are subject to $B_1$ fields of different orientation. Spins constructing the high field peak experience a $B_1$ that is always perpendicular to $B_0$, which is true of spins to the side of the wire. Due to strain these spins see a different QI, which would explain such a shift in the resonance field. A quantitative comparison between theory and experiment is the subject of ongoing work.

\section{Experimental determination of the Signal to Noise Ratio}

\begin{figure}[t]
  \centering
  \includegraphics[width=160mm]{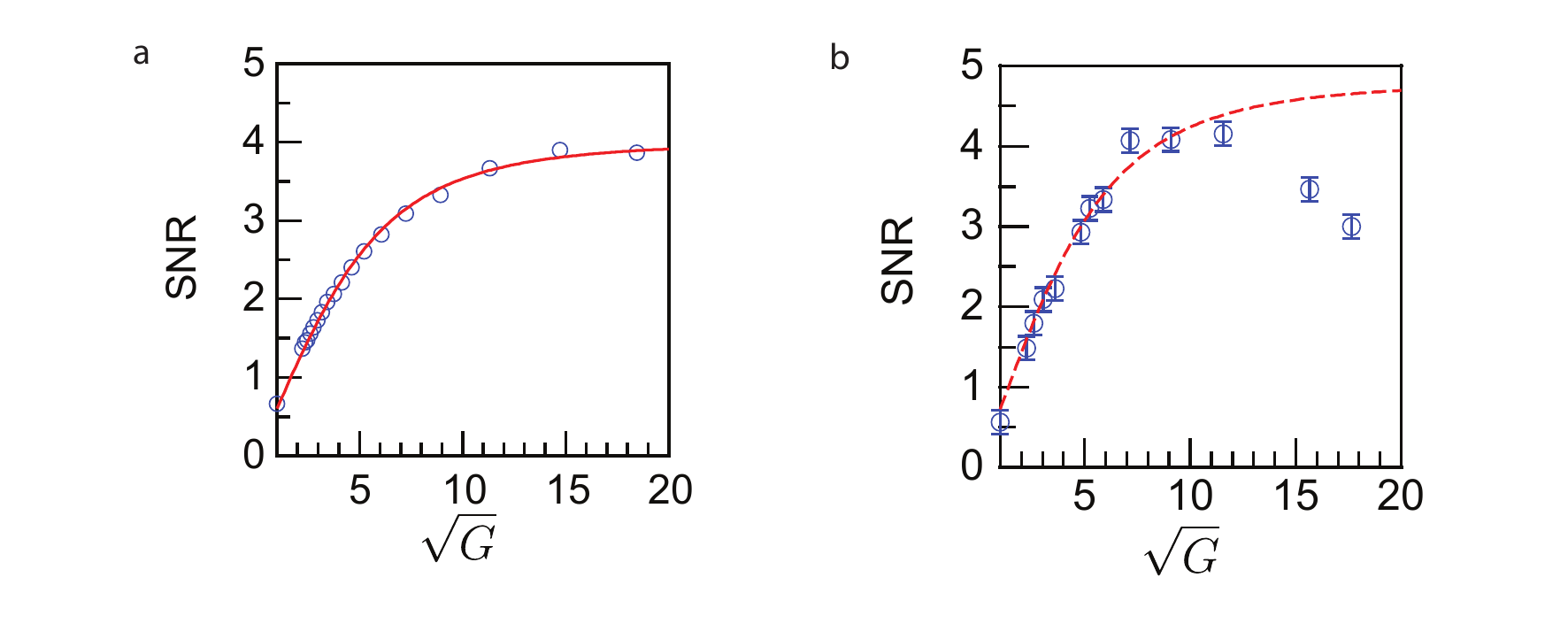}
  \caption{(a) Amplitude JPA signal-to-noise ratio $SNR=\sqrt{P_{out}/P_{noise}}$ as a function of $\sqrt{G}$ for an input microwave tone, measured with a Spectrum Analyzer, showing a six-fold improvement with the JPA on. The JPA is operated in the non-degenerate mode. Open blue circles are experimental data, red curve is a fit as explained in the text. (b) Echo signal-to-noise ratio (in amplitude) as a function of $\sqrt{G}$ (blue open circles), with error bars estimated as explained below. Red dashed line shows the amplifier $SNR(\sqrt{G})$ curve, rescaled to these data, showing that the spin-echo signal-to-noise is increased as expected from the JPA, until saturation.}
		\label{figSNR}
	\end{figure}
	
We now discuss the experimental determination of the SNR. We first characterize the signal-to-noise ratio improvement brought by the JPA itself over the following HEMT amplifier cooled at $4$\,K. A continuous microwave signal at frequency $\omega$ is sent directly on the JPA and its output spectrum $P_{out}$ is measured with a spectrum analyzer for various JPA pump power settings in the non-degenerate mode. The JPA gain in the different setting may then be computed by $G=P_{out}/ P_{out}(\text{JPA off}) $. The same experiment is then repeated without any input signal so as to obtain $P_{noise}$, the noise power in the measurement bandwidth of $100\,$kHz. The amplitude signal-to-noise ratio is then evaluated by $SNR = \sqrt{P_{out}/P_{noise}}$ and is shown in Fig.~\ref{figSNR}a. 

It follows the expected dependence $SNR \propto \sqrt{G/((G-1) n + n_{syst})}$, $n$ being the total noise photon number with the JPA ON as defined in the main text, and $n_{syst}$ being the number of noise photons added when the JPA is off. This yields a ratio $n_{syst}/n = 36$ as observed in similar setups~\cite{Castellanos.NaturePhys.4.929(2008)}, indicating $n_{syst} \approx 36$ and $n_{amp}+n_{eq} \approx 1$ therefore approaching the quantum limit, with equal contributions from $n_{eq}$ and $n_{amp}$ (while $n_s \simeq 0.01 n$ has a negligible contribution as verified experimentally). 

We then study the SNR of the spin-echo at $B_0 = 51.8$\,Gs for various JPA gains, using homodyne demodulation. For that we first choose the local oscillator phase such that the echo is entirely on one quadrature ($I$). We then average $10$ spin-echo signals yielding time-traces $\bar{I}(t)$ from which we compute the integrated echo amplitude $\mathcal{S}=\frac{1}{T_{echo}}\int_{0}^{T_{echo}}\bar{I}(t) dt$; the noise is then obtained as $\mathcal{N}=\sqrt{\frac{1}{T_{echo}}\int_{0}^{T_{echo}}\bar{I^2}(t) dt}$ when the microwave pulses are off. The noise being determined with an averaging of $500$ traces, the statistical uncertainty on the SNR comes from the signal, so that the absolute uncertainty $\epsilon_{SNR}=\epsilon_{\mathcal{S}}/\mathcal{N} = 1/\sqrt{10}$ since signal traces were averaged $10$ times. As shown in Fig.~\ref{figSNR}, the resulting $SNR=\mathcal{S}/\mathcal{N}$ follows the same dependence as already obtained for the JPA itself, which shows that with the JPA on, the detected spin-echo reaches the quantum limit of sensitivity, with a gain in sensitivity by a factor $\times 7$ compared to JPA OFF. 

It is possible to further improve the SNR by using the JPA in its degenerate mode, by setting $\omega_p = 2 \omega_0$ and setting the relative phase of the pump tone such that the amplified quadrature is $I$. This increases the gain by $6$\,dB while only increasing the noise power by $3$\,dB. The expected increase of SNR by $\sqrt{2}$ is indeed approximately observed, yielding a absolute signal-to-noise ratio of $7$, a factor $\times 11$ larger than with the JPA OFF as shown in Fig.3 of the main text.

\section{Numerical simulations}

The goal of this section is to detail the numerical fits of the absorption  and the spin-echo sequences shown in figure 3 of the main text. From this simulation, we extract the number of spins probed in a single spin-echo sequence as well as the measured spin concentration. In order to reproduce the spin dynamics, the inhomogeneity in both spin frequency and coupling strength is taken into account by dividing the ensemble into a sufficiently large set of homogeneous sub-ensembles and integrating the equations of motion for the resonator field and the spin components of all of the sub-ensembles using the model already described in\cite{Julsgaard.PhysRevA.85.013844(2012),Julsgaard.PhysRevLett.110.250503,Grezes.PHYSREVX4.021049(2014)}.

\subsection*{Model}
Consider an ensemble of $N$ spin-1/2 particles of frequency $\omega_j$. Each spin couples to the resonator field (described by creation and annihilation operators $\adagc$ and $\ac$) with a coupling constant $g_j$ and a Jaynes-Cummings interaction (see Eq.4 above). The total system Hamiltonian is then 

\begin{equation}
  \H /\hbar = \omega_0 \adagc\ac + \frac{1}{2}\sum_{j=1}^N\omega_j\pauli_z^{(j)}
      + i \sqrt{\kappa_1/2}(\beta\adagc - \beta^*\ac)
    +  \sum_{j=1}^N (g_j^* \pauli_+^{(j)}\ac + g_j \pauli_-^{(j)}\adagc),
\label{eq:Hamiltonian_working}
\end{equation}

with $\hat{\sigma}_k^{(j)}$ the Pauli operators of spin $j$ for $k = \{+,-,z\}$, and $\beta$ the amplitude of the microwave field driving the cavity input in the laboratory frame. The equations of motion are then integrated under the Markov approximation to incorporate the effect of resonator leakage and spin decoherence \cite{Grezes.PHYSREVX4.021049(2014)}. This numerical simulation yields the dynamical evolution of the mean values of the resonator field quadratures as well as of the spin operators.

The inhomogeneity in both spin resonance frequencies and coupling strengths is taken into account by dividing the entire inhomogeneous ensemble into $M$ homogeneous sub-ensembles, $\mathcal{M}_1, \mathcal{M}_2, \ldots, \mathcal{M}_M$, each of them describing  spins having an identical frequency $\omega_m$ and coupling to the cavity field $g_m$. For a sub-ensemble $m$, we define the total number of spins as $N_m$ and the three spin collective operators as $\S_i^{(m)}=\negthickspace\negthickspace\sum_{j\in \mathcal{M}_m}\pauli_i^{(j)}$, with $i\in \lbrace x,y,z \rbrace$.

Spin decoherence is treated by including a spin dephasing rate $\gammaperp= 1 / T_2$ and a spin energy decay rate $\gammapar = 1/T_1$. We use the experimentally measured coherence time $T_2=9$\,ms (see Fig.2 of the main text). The energy relaxation rate on the other hand is dominated by Purcell relaxation through the cavity, meaning that $T_1$ is longer for spins detuned from the cavity than for spins perfectly at resonance as will be discussed in later work. This is captured by defining for each ensemble $\gamma_{\parallel}^{(m)}$ as $\kappa \frac{g^{2}_{m}}{\Delta_{m}^2+\frac{\kappa ^2}{4}}$, with $\Delta_{m}=\omega_m-\omega_0$. Note that the relaxation time shown in Figure 2 $T_1=0.4\,$s was taken with a very narrow bandwidth pulse so as to obtain only the contribution from the spins that are on resonance with the cavity.

This leads us to introduce for each sub-ensemble an effective initial polarisation $S_z^{(m)} (t=0)$. Indeed, every experimental sequence is repeated several times at rate $\gamma_{rep}$ ($\gamma^{-1}_{rep} \approx $ 3 to 10s) and the results are then averaged. This waiting time $\gamma^{-1}_{rep}$  is long enough compared to $T_1$ to be neglected for spins at resonance, however detuned spins have a longer $T_1$ and thus do not fully relax between two consecutive sequences, contributing less to the signal than spins at resonance. To take into account this effect, we define an effective initial polarisation $S_z^{(m)} (t=0)$ for a sub-ensemble depending on its relaxation time:
\begin{align}
\label{eq:effecP}
S_z^{(m)} (t=0) = - N_m \times ( 1 - e^{- \gamma_{||(m)}/ \gamma_{rep}  } )  
\end{align}

The first step before performing the simulations is to determine in the context of the experiment the size $N_m$ of each sub-ensemble, which requires knowledge of the distribution of coupling constants and resonance frequency within the spin-ensemble.

\subsection*{Determining the coupling constant distribution}

\begin{figure}[!htbp]
  \centering
  \includegraphics[width=180mm]{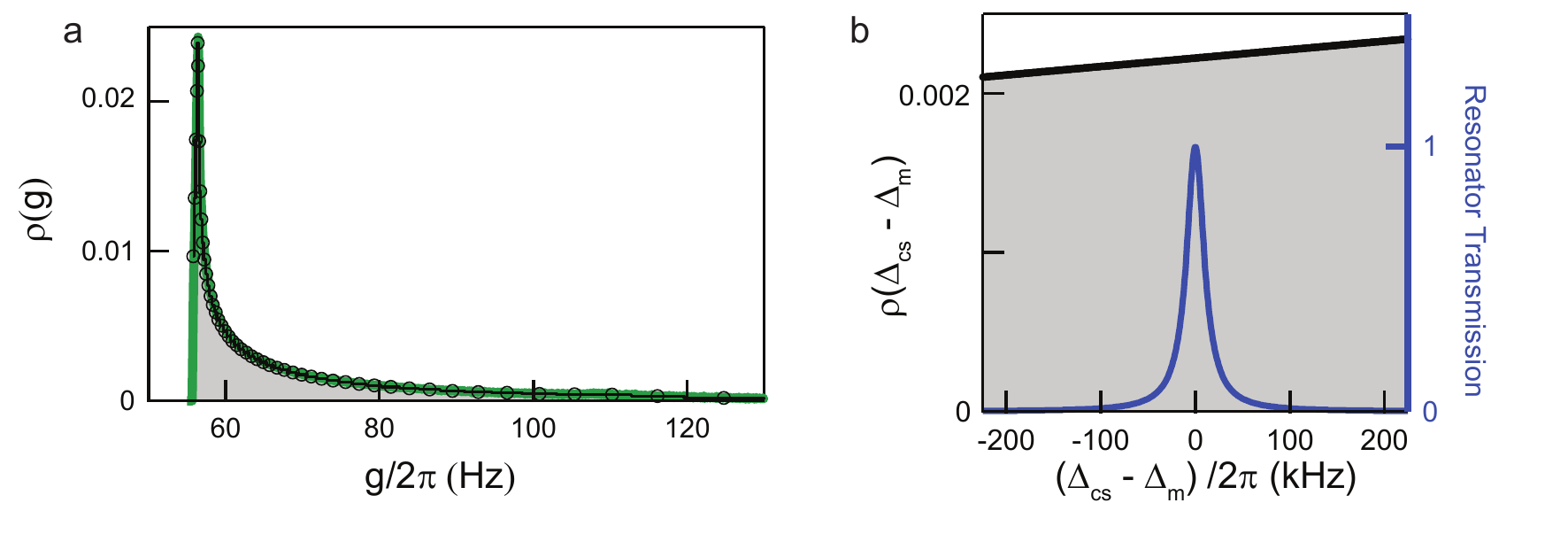}
  \caption{(a) Coupling strength distribution extracted from magnetic field simulation for $|x|<2.5\,\mu$m, with $Z_0=44\ohm$, weighted by spin concentration and normalized to unity (green line). The black circles show the discrete distribution used in the simulation, with $M_g=50$. (b) Tilted square distribution used in the simulation, with $M_{bins}=450$ (black line). The resonator transmission is plotted for comparison (blue line) }
		\label{fig3Supp}
\end{figure}

From the simulation of the vacuum fluctuations of the magnetic field shown in Figure \ref{fig2Supp}, one can compute the coupling constant distribution using Eq.~\ref{eq:g}. The measurements which we want to simulate were performed with the magnetic field $B_0$ aligned along the wire in the $\vec{z}$ direction, on the low-field peak of the structure shown on Figure 2, at $5.18\,$mT, attributed to spins lying under the wire. As a consequence, we compute the distribution only for this subset of spins, imposing $|x|<2.5 \mu m$ and $\theta=0$ in the formula. The inhomogeneous implantation profile of the spins is taken into account by appropriately weighing the coupling constant distribution. In order to normalize the distribution we take $\sum \rho(g_m)=1$. As shown in Fig.~\ref{fig3Supp}a, this yields a very asymmetric distribution sharply peaked around $g/2\pi = 56\,$Hz.

\subsection*{Determining the spin frequency distribution}

The cavity linewidth ($20\,$kHz) is two orders of magnitude smaller than the spin linewidth $3.25 \,$MHz. In order to avoid numerical errors, we use a sampling of 1 bin per $1\,$kHz, over a range of $450\,$kHz. The zero order approximation would be to assume a square spin frequency distribution, nevertheless we introduce a tilted square distribution to take into account more precisely the shape of the line, Figure \ref{fig3Supp}. The relative slope is derived from the observed spin-echo signal. At $51.8\,$mT, this yields a tilt of $ 10\%$ on the chosen range.

\subsection*{Evaluating the number of spins}

\begin{figure}[!htbp]
  \centering
  \includegraphics[width=170mm]{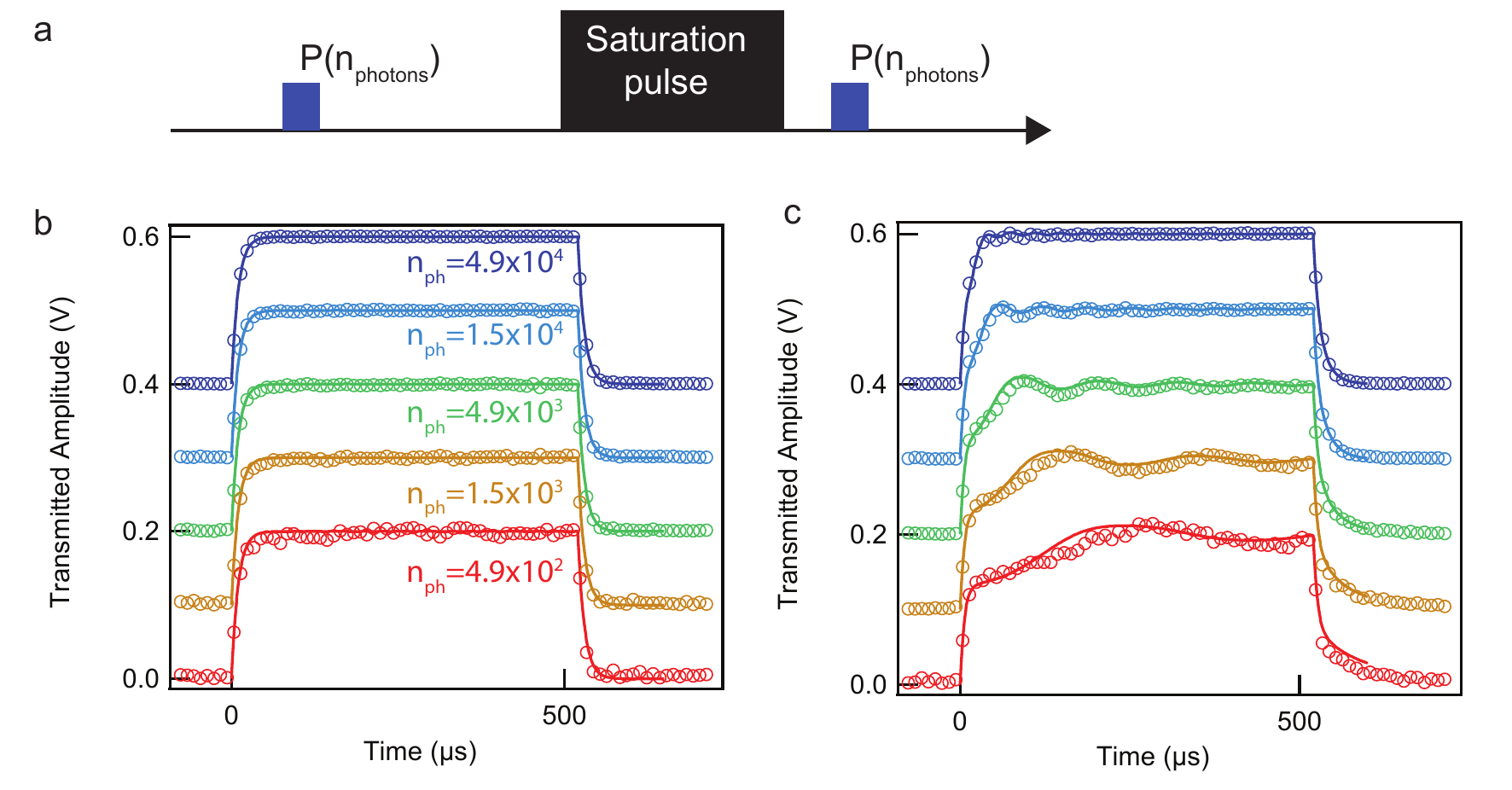}
  \caption{(a) Absorption sequence consisting of a first 500-us-long pulse at power P(nphotons), followed by a strong saturating microwave pulse immediately followed by a second 500-us-long pulse at same power P(nphotons). (b) Saturated pulses taken with average number of photons $n_{ph}$ for the intra-cavity field, rescaled to same amplitude with an additional offset and averaged 1000 times. Open circles : data, solid lines : fit. (c) Absorbed pulses, for same $n_{ph}$, rescaled thanks to saturated curves and averaged 1000 times. Open circles : experimental data, solid lines: fit}
		\label{fig4Supp}
\end{figure}

In order to determine the absolute scaling of the spin distributions, we measure the time-dependent absorption of the spins, as shown in Fig.~\ref{fig4Supp}. A first 500 $\mu$s-long pulse $P$ of power $P_{in}$ is sent to the unsaturated spins, leading to the absorption of the signal (Fig.~\ref{fig4Supp}c) whereas a second pulse of same power $P_{in}$ sent immediately after a strong microwave pulse whose role is to saturate the spins shows only the cavity dynamics (subset b). The sequence is repeated 1000 times with a repetition time $\gamma_{rep}^{-1}=3\,$s$=10 T_1$ to let the spins relax back to their ground state in-between the experimental sequences.

The transmitted pulse $P$ shows two prominent features : the Rabi oscillation transients at the beginning, which are characterized by an oscillation frequency $\Omega_R$, a decay time and an initial amplitude, and the free-induction decay (FID) of the spins which gives rise to the emission of a microwave signal even after the cavity field has decayed. 

To simulate such a sequence, the coupling distribution and the spin frequency distribution detailed above have been scaled with a total number of spins $N_{tot}=\sum\limits_{m} N_m$. The cavity parameters $\kappa_1$, $\kappa_2$, $\kappa_L$, and $\omega_c$ as well as spin relaxation rates $\gammaperp$ and $\gammapar$ are experimentally determined as explained earlier. The simulation of the saturated pulses transmission is in quantitative agreement with the data, without adjustable parameter (Figure \ref{fig4Supp}b). 

The simulated absorbed pulses are calculated with the same cavity parameters and an overall distribution scaling factor $N_{tot}=2 \times 10^5$. This parameter gives a good agreement with the FID part of the signal as well as with the Rabi oscillations amplitude. To obtain the right Rabi oscillation frequency, we simply need to scale by a factor $\eta = 1.1$ the power at the cavity input, compared to the input power estimated from the cables and filters attenuation. With only those two adjustable parameters, we are able to reproduce quantitatively the spin absorption (Figure \ref{fig4Supp}c), which is a first validation for using this model as an evaluation of the number of spins contributing to the signal.

This approach neglects one aspect of the experiment. In our sample, the inhomogeneous broadening of the line is caused by strain due to the presence of the aluminium wire. One can thus expect a correlation between the frequency of a given spin and its coupling to the resonator. Taking this correlation between coupling constant and spin frequency into account quantitatively would however require a microscopic modelling for the strain which is not available so fair. The excellent overall agreement between theory and measurements indicates that the error is minor, although this approximation may account for the small discrepancy between simulated and experimental Rabi oscillations that can be noted for low powers in Figure 4c.

\subsection*{Evaluating the number of spins contributing to the spin-echo signal}

Having as explained above calibrated the absolute scale of the spin distribution, we now evaluate the number of spins involved in a spin-echo by simulating a full Hahn echo sequence, keeping exactly the same parameters for the spin distributions. The input power of the simulated $\pi / 2$ and $\pi$ pulses are calibrated by simulating Rabi oscillations. We find that in the simulation the $\pi$ pulse power is only $1\,\textsf{dB}$ away from the experimental one, which further confirms the validity of our model.

The spin echo sequence was acquired with the JPA off, in order to avoid its saturation by the drive pulses which would distort them. The output amplitude is scaled by comparing the theoretical and experimental decay of the two excitation pulses; with only this adjustment factor the simulated echo is found to be in quantitative agreement with the experimental data as shown in Fig.3 of the main text.

\begin{figure}[!htbp]
  \centering
  \includegraphics{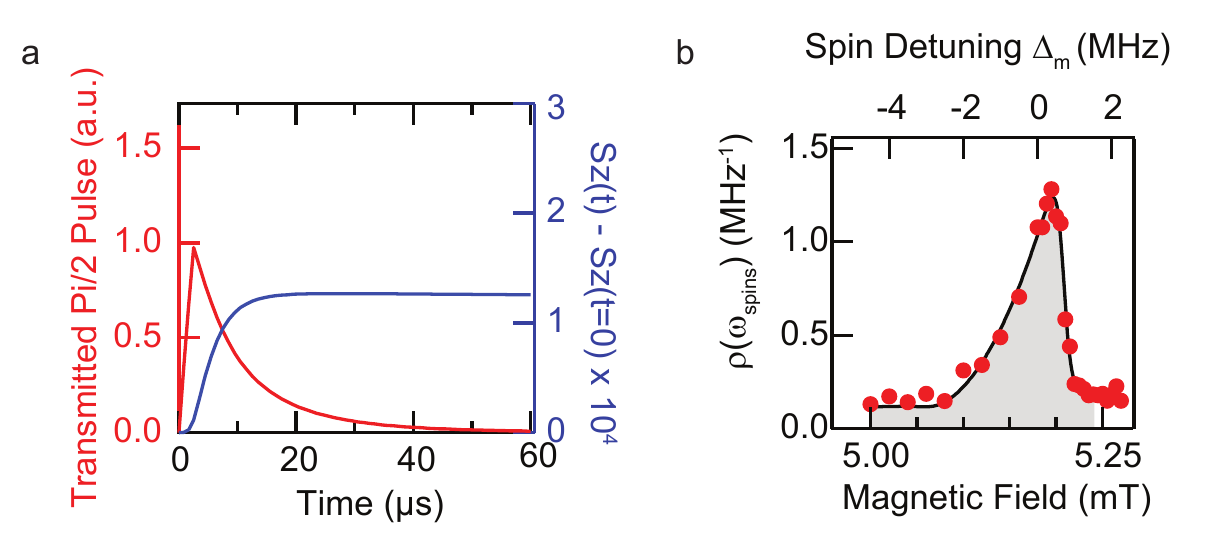}
  \caption{(a) Red: Simulated $\pi/2$ excitation pulse. Blue: $\langle S_z(t >\pi/2)  \rangle - \langle S_z(t=0)  \rangle$, allowing to extract $1.2 \times 10^4$ excited spins during this excitation pulse. (b) Spin density profile (red circles) $\propto A_e \times A_{\pi}$ (defined in the main text) rescaled to $\int \rho(\omega_{spins}) d\omega_{spins} = 1 $ (grey filled area).}
		\label{fig5Supp}
\end{figure}

To evaluate the number of spins excited during the spin-echo sequence, we extract from the simulation the time-dependent mean spin polarization $\langle S_z \rangle$, as shown in Fig.~\ref{fig5Supp}a. We consider more particularly that the quantity $\langle S_z(t >\pi/2)  \rangle - \langle S_z(t=0)  \rangle$ is a direct estimate of the number of spins excited by an Hahn-echo sequence. After the exponential decay of the $\pi/2$ excitation pulse, this value increases by $1.2 \times 10^4$. We thus come to the conclusion that $1.2 \times 10^4$ spins participate to the echo shown in Fig.3 of the main text, which is detected with a $SNR=7$, which yields the sensitivity reported in the main text.

\subsection*{Evaluating the concentration of bismuth donor spins}

Thanks to the determination of the spin density in absolute scale, we can compare the estimated number of spins to the known number of implanted atoms, to check that the two are consistent. We can scale the spin density $\rho(\omega_s) \propto A_e \times A_{\pi}$ thanks to the absorption simulation. By integrating the lower field peak, Figure \ref{fig5Supp}b, we find a total number of spins of $1.07 \times 10^6$ contributing to the absorption. As Bismuth donors have a nuclear spin of $9/2$, this represents only one-tenth of the total amount of spins. Also as this peak has been identified as signal emitted only by spins under the wire, the amount of spins is diluted in a surface $S_{wire}= L_{wire} \times W_{wire} = 720 \mu m \times 5\mu m = 3.6\times 10^{-5} \,cm^2$, leading to an experimental surface  concentration $[ Bi ]_{Exp} = 10 \times 1.07 \times 10^6/( 3.6 \times 10^-5 )= 2.97 \times 10^{11} \, cm ^{-2}$, a number which is only a factor 3 lower than the surface concentration extracted by SIMS measurement $[ Bi ] _{SIMS} = 9.45 \times 10^{11}\, cm ^{-2}$ . 

This ratio of 30\% can be explained by two factors: first that the activation of bismuth atoms i.e. the migration of bismuth atoms from interstitial implantation site to substitutional site by rapid annealing is not total. This factor has been evaluated in \cite{Weis.APL.100.172104(2012)} to be 60\% by electrical measurement. The additional factor 2 with our experiment may be due to a fraction of bismuth atoms being in an ionized state and thus not contributing to the ESR signal.

\subsection*{Approximate analytical formula for the expected spin-echo signal-to-noise ratio}

We now derive an approximate analytical formula for the pulsed ESR spectrometer sensitivity, which is given in the main text. The goal is to provide an analytical estimate for the
experimentally obtained sensitivity and thus identify how the sensitivity scales with the physical parameters.

Compared to the model used in section IV, we consider here for simplicity that all spins have equal coupling constant $g$ to the resonator. As a result in the frame rotating at the
resonator frequency $\omega_0$, the Hamiltonian is:
\begin{equation}
\label{eq:Hamiltonian}
  \H /\hbar = \sum_j\left[\frac{\Delta_j}{2}\pauli^{(j)}_z + g(\ac^\dagger
  \pauli^{(j)}_- + \ac\pauli^{(j)}_+)\right].
\end{equation}
To obtain analytical results, it is convenient to assume that the spin 
detunings $\Delta_j$ are distributed according to a Lorentzian function $f(\Delta) =
\frac{w/2\pi}{\Delta^2 + w^2/4}$, with a FWHM of $w$. More realistic distributions would only change
the final results by a factor of order unity. For simplicity again, we neglect decoherence 
of the spins. The following equations of motion then
describe the evolution of mean values:

\begin{align}
\label{eq:dac/dt}
  \frac{d \mean{\ac(t)}}{dt} & =  - \frac{\kappa}{2} \mean{\ac(t)}
    -ig\sum_j \mean{\pauli^{(j)}_-(t)}, \\
\label{eq:dpauli-/dt}
  \frac{d\mean{\pauli^{(j)}_-(t)}}{dt} &= -i\Delta_j\mean{\pauli^{(j)}_-(t)}
     + ig\mean{\pauli^{(j)}_z(t)\ac(t)}
\end{align}
We define $\S_- = \sum_j\pauli_-^{(j)}$, and formally integrate (\ref{eq:dac/dt}),
\begin{equation}
\label{eq:ac_formal_int}
  \mean{\ac(t)} = -ig\int_{-\infty}^t e^{-\frac{\kappa}{2}(t-t')}\mean{\S_-(t')}dt'.
\end{equation}
From this equation we see that $\mean{\ac}$ attains at most a value of
$\approx gN/\kappa$, and if $g$ times this value is much smaller than
typical detunings, $\Delta_j$, \textit{i.e.}, the
cooperativity parameter $C = \frac{4g^2N}{\kappa w} \ll 1$, the second term of
Eq.~(\ref{eq:dpauli-/dt}) can be neglected. We assume this is the case and we thus treat 
the spins as evolving freely between the pulses.

We will focus on the case of a standard Hahn-echo sequence,
$\pi/2$--$\tau$--$\pi$--$\tau$, and we assume perfect 
initial $\pi/2$  and refocusing $\pi$ pulses at times $-2\tau$ and $-\tau$, respectively. 
Prior to $t=-2\tau$, the spins are polarized with polarization $p \le 1$ along
the $-z$-direction such that $\mean{\pauli_z^{(j)}} = -p$ and
$\mean{\pauli_x^{(j)}} = \mean{\pauli_y^{(j)}} = 0$. The case $p=1$
corresponds to a perfectly polarized sample with all spins in the
ground state. When the spin ensemble is subjected to the perfect
$\pi/2$ pulse around the $x$-axis at time $t=-2\tau$, a state with
$\mean{\pauli_y^{(j)}(0)} = p$ and hence $\mean{\pauli_-^{(j)}(0)} =
-\frac{ip}{2}$ is prepared. The refocusing $\pi$
pulse at $t=-\tau$ implies that a spin echo will occur at time $t = 0$, where the free evolution, $\mean{\pauli_-^{(j)}(t)} =
-\frac{ipe^{-i\Delta_j t}}{2}$, cause all spins to be in phase. The resulting total spin  $\mean{\S_-(t)}$ can be calculated in the continuum limit, where the sum over spins is replaced by an integral over detunings,
\begin{equation}
\label{eq:Sminus_Lorentz}
  \mean{\S_-(t)} = \int f(\Delta)\frac{-ipN}{2}e^{-i\Delta t} d\Delta
  = -\frac{ipN}{2}e^{-w|t|/2}.
\end{equation}
Inserting this into Eq.~(\ref{eq:ac_formal_int}) leads to the
following time dependence of the resonator field:
\begin{equation}
\label{eq:ac_Lorentz}
  \mean{\ac(t)} = -\frac{gpN}{\kappa+w}\times\left\{
      \begin{matrix}
        e^{\frac{wt}{2}} & \quad t<0, \\
       \frac{\kappa+w}{\kappa-w}e^{-\frac{wt}{2}} -
       \frac{2w}{\kappa-w}e^{-\frac{\kappa t}{2}} & \quad t>0.
      \end{matrix}
      \right.
\end{equation}
Now, as indicated in Fig.~(1) of the main text, this resonator field
is coupled with rate $\kappa_2$ into a Josephson parametric amplifier
(JPA). The input field, $\ain$, to this amplifier is related to the
resonator field, $\ac$, by $\mean{\ain(t)} =
\sqrt{\kappa_2}\mean{\ac(t)}$ and is normalized such that
$\mean{\ain^{\dagger}(t)\ain(t)}$ represents the number of microwave
photons per time incident on the amplifier. We shall also define the
input field quadrature variables by $\Xin(t) = \frac{\ain(t) +
  \ain^\dagger(t)}{2}$ and $\Yin(t) = \frac{-i(\ain(t) -
  \ain^\dagger(t))}{2}$. Since $\mean{\ac(t)}$ is real-valued in our
calculations, see Eq.~(\ref{eq:ac_Lorentz}), the mean signal is
carried by the $\hat{X}$-quadrature.

So far we have only considered mean values of the spins and the cavity field, and we shall apply an operator 
description of the amplification stage to assess the noise on the measurement signal. 
To this end we define
single modes of the propagating field by $\ain = \int \ain(t) u(t)dt$, where
$u(t)$ is a mode function chosen to be real valued and fulfilling
$\int [u(t)]^2 dt = 1$. In this case we have $[\ain,\ain^\dagger] =
\iint u(t)u(t')[\ain(t),\ain^\dagger(t')]dt dt' = 1$ due to the free field commutator 
relations $[\ain(t), \ain(t')]= \delta(t-t')$. The corresponding single mode quadrature variables,
$\Xin = \int\Xin(t)u(t)dt$ and $\Yin = \int \Yin(t)u(t)dt$, fulfill
$[\Xin,\Yin] = \frac{i}{2}$, and the minimum
uncertainty state, obtained for the vacuum state or coherent states of
the field, must obey $\langle \Delta\Xin^2 \rangle= \langle \Delta\Yin^2 \rangle= \frac{1}{4}$. 

Noise in the propagating field is conveniently characterized by its dimensionless power spectrum $S(\omega)/\hbar \omega = \langle \Delta\Xin^2 + \Delta\Yin^2 \rangle$. At the cavity output, noise arises from both electromagnetic equilibrium fluctuations, characterized by $\neq$, and from possible extra noise due to spontaneous emission of the spins, as will be discussed further, with a contribution $\nsp$. Our experiments being performed at temperatures such that $k T \ll \hbar \omega_0$, the electromagnetic field at equilibrium is indeed very close to its ground state so that we can safely assume that $\neq = 1/2$. In total we get $\langle \Delta \Xin^2 \rangle = \frac{1}{2}(\nsp + \frac{1}{2})$.

When the signal pulse, emitted from the resonator, is transmitted through the amplifier, its power is increased by the gain $G$. However the amplification process itself can add extra noise to the output field, characterized by a dimensionless power density $\namp$, and further degrade the signal-to-noise ratio. Following Caves~\cite{Caves.PhysRevD.26.1817(1982)}, two cases should then be envisioned to describe the statistics of the field at the amplifier output. If the amplifier is in the so-called non-degenerate mode, its single mode output is described by the field annihilation operator:
\begin{equation}
\label{inout}
  \aout = \sqrt{G}\ \ain + \sqrt{G-1}\ \bid^\dagger.
\end{equation}
To ensure the bosonic commutator relation of the amplified
signal operators, an idler mode operator, obeying
$[\bid,\bid^\dagger]=1$, must be included in the amplifier relation~\cite{Caves.PhysRevD.26.1817(1982)}. It should also be noted that the gain $G$ is generally a
function of frequency, which may distort the temporal shape of an
amplified pulse. However, the measured 1.7 MHz wide gain profile, see
Fig.~1(f) of the main text, supports that we can assume a constant
gain across the bandwidth of the signal and hence for the 
single mode defined by $u(t)$.

We introduce quadrature operators for the output and idler modes in a
similar manner as for the input field, and the mean value of the input
field is simply amplified as $\mean{\Xout} =
\sqrt{G}\mean{\Xin}$. However, the noise in the output has
contributions from both the signal and the idler mode,
\begin{equation} \label{noise}
  \langle \Delta \Xout^2 \rangle  = G \langle \Delta \Xin^2 \rangle + (G-1) \langle \Delta \Xid^2 \rangle.
\end{equation}
Assuming the idler mode thermalized at a temperature $T$ yields $\langle \Delta \Xid^2 \rangle = \frac{1}{4}(1 + 2 \bar{n})$, $\bar{n} = 1/ (\mathrm{e}^{\frac{\hbar \omega_0}{k T} } - 1)$ being the
mean thermal photon number. At high temperatures $\frac{\kB T}{\hbar \omega_0}\gg
1$ so that $\overline{n}\simeq \frac{\kB T}{\hbar \omega_0}$, the thermal state of the idler yields an overwhelming
contribution $(G-1)\frac{\namp}{2}$ to the readout noise on the $X$ quadrature, with $\namp = \bar{n}$. This is in particular the case in our experiments when the JPA is turned off and the signal is exclusively amplified by the HEMT amplifier at the $4$K stage, for which $\namp \sim 50$. When the JPA is on, this contribution is minimized since amplification is now carried on at a temperature $T$ such that $\kB T \ll
\hbar\omega_0$, with an amplifier that reaches the quantum limit. In these conditions, $\namp = 1/2$.

In total, we find that the amplification obeys the following relation for the signal-to-noise ratio :
\begin{equation} \label{signoise}
  \frac{\mean{\Xout}^2/\Delta \Xout^2}{\mean{\Xin}^2/\Delta\Xin^2}=
    \frac{G (\neq + \nsp)}{G (\neq + \nsp) + (G-1)\namp}.
\end{equation}
In the special case that~$\namp = \neq = \frac{1}{2}$ and assuming $\nsp = 0$, this equals
$\frac{G}{2G-1}$, which yields, in the limit of large $G$, the well
known factor of two (i.e.~3 dB) reduction in squared signal-to-noise
by phase insensitive amplification with no excess noise. After
amplification by the JPA, the signal level is sufficient that
amplification and homodyne demodulation do not further degrade the
signal-to-noise. In our analytical estimate of the ESR sensitivity, we 
shall proceed with the assumption that the temperature 
is sufficiently low, the gain is sufficiently high, and the excess spin noise is negligible
(we shall return briefly to an assessment of this assumption). 

The other case to consider is the one where the amplifier is phase-sensitive, with quadrature-dependent gains $G_{X,Y}$. This ensures
the correct output commutator relations without requiring the addition
of idler mode noise as explained in~\cite{Caves.PhysRevD.26.1817(1982)}. An ideal
amplifier at the quantum limit may then verify $G_X = G_Y^{-1} \gg 1$
while implementing noiseless amplification of one of the two
quadratures, implying that in this case $\langle \Delta \Xout^2 \rangle = n/2$, with
$n=n_{\mathrm{eq},X}+n_{\mathrm{amp},X}$ and $n_{\mathrm{amp},X} =
0$.

Summing up the discussion, one sees that the noise on quadrature $X$ referred to the JPA input can be written as $\langle \Delta \Xout^2 \rangle = n/2$, with $n = \neq + \namp + \nsp$. In our experiment, $\neq \sim 1/2$, whereas $\namp \sim 50$ if the JPA is off, $\namp \sim 1/2$ if it is operated in the non-degenerate mode, and $\namp \sim 0$ if it is operated in degenerate mode. 

Eqs.~(\ref{noise}, \ref{signoise}) account for the noise and
signal-to-noise for measurements of the continuous output amplitude
signal weighted with $u(t)$. The optimal choice for $u(t)$ is the one that maximizes
the weighted signal $\mean{\Xout}$ without altering the noise
(assuming uniform broad band noise of the idler mode $\bid$). In an experiment one may
choose $u(t)$ as the measured shape of the emitted pulse averaged over many
experimental runs. As shown by the red curve in the inset of Fig.~3(c) in the main text, we can obtain the same shape
by a numerical calculation of the spin dynamics. Since we are here interested in analytical estimates, we choose the result given by 
$u(t) = \mean{\Xin(t)}/\mean{\Xin}$, where $\mean{\Xin(t)} =
\frac{1}{2}\mean{\ain(t)+\ain^\dagger(t)} = \frac{\sqrt{\kappa_2}}{2}
\mean{\ac(t)+\ac^\dagger(t)} = \sqrt{\kappa_2}\mean{\ac(t)}$, with an explicit expression given in  Eq.(\ref{eq:ac_Lorentz}). To obtain the correct normalization  $\int[u(t)]^2dt = 1$, we
calculate the squared signal as:
\begin{equation}
  \mean{\Xin}^2 = \int \mean{\Xin(t)}^2 dt =
      \frac{2g^2p^2N^2\kappa_2(\kappa+2w)}{(\kappa+w)^2 w\kappa}.
\end{equation}
The output signal-to-noise ratio then reads:
\begin{equation}
  \frac{|\mean{\Xout}|}{\Delta\Xout} =
  \frac{2gpN}{\kappa+w} \sqrt{\frac{1}{n}} \sqrt{\frac{\kappa_2(\kappa+2w)}{w\kappa}}.
\end{equation}
The minimum number of detectable spins, which defines the ESR spectrometer sensitivity, is thus
\begin{equation}
  \Nmin =
  \frac{\kappa+w}{2gp}\sqrt{\frac{n w\kappa}{\kappa_2(\kappa+2w)}}
  \rightarrow
  \frac{\kappa}{2gp} \sqrt{\frac{n w}{\kappa_2}}.
\end{equation}
The arrow indicates the limit of conventional ESR operation, $\kappa
\gg w$. Introducing the echo duration $T_E = w^{-1}$ as seen from Eq.~\ref{eq:ac_Lorentz}, and taking the case of a critically coupled resonator for which $\kappa = 2\kappa_2$, one obtains $\Nmin = \frac{1}{g p} \sqrt{\frac{\kappa n}{T_E}}$, which is the formula found in the main text in the experimentally relevant case $p=1$. For our parameters, this yields $N_{min} = 400$.

This estimate can be further refined to take into account the fact that the actual width
$\Gamma/2\pi \approx 1$ MHz of the spin frequency distribution is much
larger than $\kappa$. In this case, the $\pi/2$ and $\pi$ pulses
excite a subset of spins, and according to the numerical simulations,
this subset has a Lorentzian profile with $w/2\pi \approx 25$ kHz,
which is very close to $\kappa/2\pi \approx 22$ kHz. Hence, in the
above expression (to the left of the arrow) one can take $\kappa \approx w
\approx 2\kappa_2$ for critical coupling, which yields $\Nmin \approx
\sqrt{\frac{2}{3}}\frac{\kappa \sqrt{n}}{g} \approx 3\cdot 10^2$, using
$g/2\pi\approx 55$ Hz and $p \approx 1$. This refinement should however not be considered too seriously when compared to the experimental data, given the other approximations that were made, such as perfect $\pi$ pulse, Lorentzian line profile, and optimal weighing function $u(t)$.

The signal is further enhanced, with no increase in the noise, when
accumulated over the CPMG echoes, i.e., by choosing the corresponding
multi-peaked $u(t)$, cf. Fig.~4(a) of the main text. In the analysis of the experiments, we assumed a weighting of the 
signals by tophat pulses of equal weight, and we, indeed, observed an improved signal-to-noise over the single pulse
analysis. Let us here estimate the theoretical limitations of the CPMG echo spectroscopy, assuming a gradual, exponential 
reduction of the echo amplitudes. With $m$ pulses
in total, we may define the corresponding normalized mode function
$u^{(m)}(t)$ implicitly as:
\begin{equation}
  u^{(m)}(t) = \frac{\sum_{j=0}^{m-1}\mean{\Xin^{(1)}(t - jT)}e^{-jT/T_{CPMG}}}
    {\mean{\Xin^{(m)}}},
\end{equation}
where $\mean{\Xin^{(1)}(t)}$ is the output quadrature for a single
$\pi$ pulse and we assume that the echoes are non-overlapping and
simply repeated with a period of $T$ but damped by the rate
$T_{CPMG}^{-1}$. This $T_{CPMG}$ may include any experimentally determined damping
effects, such as non-ideal $\pi$ pulses causing a degradation of the
signal. As in the above analysis, the normalization, $\int [u^{(m)}(t)]^2dt = 1$ is
automatically accounted for when calculating the squared signal as:
\begin{equation}
  \mean{\Xin^{(m)}}^2 = \int\sum_{j=0}^{m-1}\mean{\Xin^{(1)}(t - jT)}^2 e^{-2jT/T_{CPMG}}dt
  = \mean{\Xin^{(1)}}^2\frac{1 - e^{-2mT/T_{CPMG}}}{1-e^{-2T/T_{CPMG}}}.
\end{equation}
The use of multiple pulses in the CPMG protocol only makes sense if
$T\ll T_{CPMG}$, and the denominator in the above expression can be
approximated as $2T/T_{CPMG}$. Hence, the signal-to-noise ratio, SNR$_m$
for $m$ pulses becomes:
\begin{equation}
  \frac{\mathrm{SNR}_m}{\mathrm{SNR}_1} = \sqrt{\frac{T_{CPMG}}{2T}(1 - e^{-2mT/T_{CPMG}})},
\end{equation}
which behaves as  $\sqrt{m}$ for a small number of pulses with $2 m T
\ll T_{CPMG}$ and saturates at $\sqrt{T_{CPMG}/2T}$ when $2mT \gg T_{CPMG}$.

Let us finally return to our assumption that the spin noise is negligible.
To assess this issue we have performed a numerical calculation similar to the one shown in
Fig.~3(c) in the main text, but including also the quantum noise using
the methods of~\cite{Julsgaard.PhysRevA.85.013844(2012)}. For an effective number of
spins $N \approx 1.2\cdot 10^4$, this
calculation, indeed, shows an excess noise of $\approx 30$ \%, and
we find that this noise is proportional to the number of spins
involved. Since the excess noise is calculated for the situation where the
signal-to-noise ratio is $7\pm 1$, it must be seven times smaller, i.e.,
only at a few percent level when the signal-to-noise is unity. For this reason, it
does not affect the experimental assessment of $\Nmin \approx 1.7\cdot
10^3$ in the main text. We also note that with the effective number of
spins $N \approx 1.2\cdot 10^4$, the cooperativity parameter reaches
the value of $C = \frac{4g^2N}{\kappa w} \approx 0.26$. This number is also proportional to $N$ and thus
approximately seven times smaller in the case of a signal-to-noise
level of unity, thus validating our assumption, $C \ll 1$, in the analytical
estimate.

\end{document}